\documentclass[twocolumn]{aastex7}

\usepackage{amsmath}
\usepackage{multirow} 

\newcommand{\ad}[1]{{\color{black} {#1}}}
\def\ion#1#2{#1$\;${\footnotesize\rm{#2}}\relax}

\begin{document}

\title{A Massive Black Hole 0.8\,kpc from the Host Nucleus \\ Revealed by the Offset Tidal Disruption Event AT2024tvd}

\author[0000-0001-6747-8509]{Yuhan Yao}
\email{yuhanyao@berkeley.edu}
\affiliation{Miller Institute for Basic Research in Science, 468 Donner Lab, Berkeley, CA 94720, USA}
\affiliation{Department of Astronomy, University of California, Berkeley, CA 94720-3411, USA}

\author[0000-0002-7706-5668]{Ryan Chornock}
\email{chornock@berkeley.edu}
\affiliation{Department of Astronomy, University of California, Berkeley, CA 94720-3411, USA}

\author[0000-0002-4557-6682]{Charlotte Ward}
\email{charlotte.ward@princeton.edu}
\affil{Department of Astrophysical Sciences, Princeton University, Princeton, NJ 08544, USA} 

\author[0000-0002-5698-8703]{Erica Hammerstein}\email{ekhammer@berkeley.edu}
\affiliation{Department of Astronomy, University of California, Berkeley, CA 94720-3411, USA}

\author[0000-0003-0466-3779]{Itai Sfaradi}\email{itai.sfaradi@berkeley.edu}
\affiliation{Department of Astronomy, University of California, Berkeley, CA 94720-3411, USA}

\author[0000-0003-4768-7586]{Raffaella Margutti}\email{rmargutti@berkeley.edu}
\affiliation{Department of Astronomy, University of California, Berkeley, CA 94720-3411, USA}
\affiliation{Department of Physics, University of California, 366 Physics North MC 7300, Berkeley, CA 94720, USA}

\author[0000-0002-6625-6450]{Luke Zoltan Kelley}\email{lzkelley@berkeley.edu}
\affiliation{Department of Astronomy, University of California, Berkeley, CA 94720-3411, USA}

\author[0000-0002-1568-7461]{Wenbin Lu}\email{wenbinlu@berkeley.edu}
\affiliation{Department of Astronomy, University of California, Berkeley, CA 94720-3411, USA}
\affiliation{Theoretical Astrophysics Center, University of California, Berkeley, CA 94720, USA}

\author[0000-0002-7866-4531]{Chang~Liu}\email{ptg.cliu@u.northwestern.edu}
\affiliation{Department of Physics and Astronomy, Northwestern University, 2145 Sheridan Rd, Evanston, IL 60208, USA}
\affiliation{Center for Interdisciplinary Exploration and Research in Astrophysics (CIERA), Northwestern University, 1800 Sherman Ave, Evanston, IL 60201, USA}

\author[0000-0003-0733-2916]{Jacob Wise}\email{J.L.Wise@2022.ljmu.ac.uk}
\affiliation{Astrophysics Research Institute, Liverpool John Moores University, 146 Brownlow Hill, Liverpool L3 5RF, UK}

\author[0000-0003-1546-6615]{Jesper Sollerman}\email{jesper@astro.su.se}
\affiliation{The Oskar Klein Centre, Department of Astronomy, Stockholm University, AlbaNova, SE-10691, Stockholm, Sweden}


\author[0000-0002-8297-2473]{Kate D. Alexander}\email{kdalexander@arizona.edu}
\affiliation{Department of Astronomy/Steward Observatory, 933 North Cherry Avenue, Room N204, Tucson, AZ 85721-0065, USA}

\author[0000-0001-8018-5348]{Eric C. Bellm}\email{ecbellm@uw.edu}
\affiliation{DIRAC Institute, Department of Astronomy, University of Washington, 3910 15th Avenue NE, Seattle, WA 98195, USA}

\author{Andrew J. Drake}\email{ajd@astro.caltech.edu}
\affiliation{Cahill Center for Astrophysics, California Institute of Technology, MC 249-17, 1200 E California Boulevard, Pasadena, CA 91125, USA}

\author[0000-0002-4223-103X]{Christoffer Fremling}\email{fremling@caltech.edu}
\affiliation{Caltech Optical Observatories, California Institute of Technology, Pasadena, CA 91125, USA}
\affiliation{Cahill Center for Astrophysics, California Institute of Technology, MC 249-17, 1200 E California Boulevard, Pasadena, CA 91125, USA}

\author{Marat Gilfanov}\email{marat.gilfanov@gmail.com}
\affiliation{Space Research Institute, Russian Academy of Sciences, Profsoyuznaya ul. 84/32, Moscow, 117997, Russia}
\affiliation{Max-Planck-Institut f\"{u}r Astrophysik, Karl-Schwarzschild-Str. 1, D-85741 Garching, Germany}

\author[0000-0002-3168-0139]{Matthew J. Graham}\email{mjg@caltech.edu}
\affiliation{Cahill Center for Astrophysics, California Institute of Technology, MC 249-17, 1200 E California Boulevard, Pasadena, CA 91125, USA}

\author[0000-0001-5668-3507]{Steven L. Groom}\email{sgroom@ipac.caltech.edu}
\affiliation{IPAC, California Institute of Technology, 1200 E. California Blvd, Pasadena, CA 91125, USA}

\author[0000-0002-0129-806X]{K. R. Hinds}\email{K.C.Hinds@2021.ljmu.ac.uk}
\affiliation{Astrophysics Research Institute, Liverpool John Moores University, 146 Brownlow Hill, Liverpool L3 5RF, UK}

\author[0000-0001-5390-8563]{S.~R.~Kulkarni}\email{srk@astro.caltech.edu}
\affiliation{Cahill Center for Astrophysics, California Institute of Technology, MC 249-17, 1200 E California Boulevard, Pasadena, CA 91125, USA}

\author[0000-0001-9515-478X]{Adam A. Miller}\email{amiller@northwestern.edu}
\affiliation{Department of Physics and Astronomy, Northwestern University, 2145 Sheridan Rd, Evanston, IL 60208, USA}
\affiliation{Center for Interdisciplinary Exploration and Research in Astrophysics (CIERA), Northwestern University, 1800 Sherman Ave, Evanston, IL 60201, USA}
\affiliation{NSF-Simons AI Institute for the Sky (SkAI), 172 E. Chestnut St., Chicago, IL 60611, USA}

\author[0000-0003-3124-2814]{James C. A. Miller-Jones}\email{james.miller-jones@curtin.edu.au}
\affiliation{International Centre for Radio Astronomy Research Curtin University, GPO Box U1987, Perth, WA 6845, Australia}

\author[0000-0002-2555-3192]{Matt Nicholl}\email{mattnicholl1@gmail.com}
\affiliation{Astrophysics Research Centre, School of Mathematics and Physics, Queens University Belfast, Belfast BT7 1NN, UK}

\author[0000-0001-8472-1996]{Daniel A.~Perley}\email{d.a.perley@ljmu.ac.uk}
\affiliation{Astrophysics Research Institute, Liverpool John Moores University, 146 Brownlow Hill, Liverpool L3 5RF, UK}

\author[0000-0003-1227-3738]{Josiah Purdum}\email{jpurdum@caltech.edu}
\affiliation{Caltech Optical Observatories, California Institute of Technology, Pasadena, CA 91125, USA}

\author[0000-0002-7252-5485]{Vikram Ravi}\email{v.vikram.ravi@gmail.com}
\affiliation{Cahill Center for Astrophysics, California Institute of Technology, MC 249-17, 1200 E California Boulevard, Pasadena, CA 91125, USA}

\author[0000-0003-0427-8387]{R. Michael Rich}\email{rmrastro@gmail.com}
\affiliation{Department of Physics \& Astronomy, University of California Los Angeles, 430 Portola Plaza, Los Angeles, CA 90095, USA}

\author[0000-0002-5683-2389]{Nabeel Rehemtulla}\email{nabeelr@u.northwestern.edu}
\affiliation{Department of Physics and Astronomy, Northwestern University, 2145 Sheridan Rd, Evanston, IL 60208, USA}
\affiliation{Center for Interdisciplinary Exploration and Research in Astrophysics (CIERA), Northwestern University, 1800 Sherman Ave, Evanston, IL 60201, USA}
 \affiliation{NSF-Simons AI Institute for the Sky (SkAI), 172 E. Chestnut St., Chicago, IL 60611, USA}

\author[0000-0002-0387-370X]{Reed Riddle}\email{riddle@caltech.edu}
\affiliation{Caltech Optical Observatories, California Institute of Technology, Pasadena, CA 91125, USA}

\author[0000-0001-7062-9726]{Roger Smith}\email{rsmith@astro.caltech.edu}
\affiliation{Caltech Optical Observatories, California Institute of Technology, Pasadena, CA  91125, USA}

\author[0000-0003-2434-0387]{Robert Stein}\email{rdstein@umd.edu}
\affiliation{Department of Astronomy, University of Maryland, College Park, MD 20742, USA}
\affiliation{Joint Space-Science Institute, University of Maryland, College Park, MD 20742, USA} 
\affiliation{Astrophysics Science Division, NASA Goddard Space Flight Center, Mail Code 661, Greenbelt, MD 20771, USA} 

\author{Rashid Sunyaev}\email{sunyaev@mpa-garching.mpg.de}
\affiliation{Space Research Institute, Russian Academy of Sciences, Profsoyuznaya ul. 84/32, Moscow, 117997, Russia}
\affiliation{Max-Planck-Institut f\"{u}r Astrophysik, Karl-Schwarzschild-Str. 1, D-85741 Garching, Germany}

\author[0000-0002-3859-8074]{Sjoert van Velzen}\email{sjoert@strw.leidenuniv.nl}
\affiliation{Leiden Observatory, Leiden University, Postbus 9513, 2300 RA, Leiden, The Netherlands}

\author[0000-0002-9998-6732]{Avery Wold}\email{awold@ipac.caltech.edu}
\affiliation{IPAC, California Institute of Technology, 1200 E. California Blvd, Pasadena, CA 91125, USA}

\begin{abstract}

Tidal disruption events (TDEs) that are spatially offset from the nuclei of their host galaxies offer a new probe of massive black hole (MBH) wanderers, binaries, triples, and recoiling MBHs. Here we present AT2024tvd, the first off-nuclear TDE identified through optical sky surveys. High-resolution imaging with the \textit{Hubble Space Telescope} shows that AT2024tvd is $0.914\pm 0.010^{\prime\prime}$ offset from the apparent center of its host galaxy, corresponding to a projected distance of $0.808\pm 0.009$\,kpc at $z=0.045$. Chandra and VLA observations support the same conclusion for the TDE’s X-ray and radio emission. AT2024tvd exhibits typical properties of nuclear TDEs, including a persistent hot UV/optical component that peaks at $L_{\rm bb}\sim 6\times 10^{43}\,{\rm erg\,s^{-1}}$, broad hydrogen lines in its optical spectra, and delayed brightening of luminous ($L_{\rm X,peak}\sim 3\times 10^{43}\,{\rm erg\,s^{-1}}$), highly variable soft X-ray emission. The MBH mass of AT2024tvd is $10^{6\pm1}\,M_\odot$, at least 10 times lower than its host galaxy's central black hole mass ($\gtrsim 10^8\,M_\odot$). The MBH in AT2024tvd has two possible origins: a wandering MBH from the lower-mass galaxy in a minor merger during the dynamical friction phase or a recoiling MBH ejected by triple interactions. Combining AT2024tvd with two previously known off-nuclear TDEs discovered in X-rays (3XMM\,J2150 and EP240222a), which likely involve intermediate-mass black holes in satellite galaxies, we find that the parent galaxies of all three events are very massive ($\sim 10^{10.9}\,M_\odot$). This result aligns with expectations from cosmological simulations that the number of offset MBHs scales linearly with the host halo mass. 

\end{abstract}


\section{Introduction}

The hierarchical merger-driven process of galaxy assembly naturally predicts the existence of massive black hole (MBH) pairs and MBH binaries \citep{Tremmel2018_mnras, Ricarte2021_wanderers_origin}, as almost every bulge-dominant galaxy harbors a central MBH \citep{Kormendy2013}. The journey from galactic scales to the eventual MBH merger involves multiple processes operating across a range of spatial scales \citep{Begelman1980}.

On large scales ($\sim{\rm kpc}$), dynamical friction (DF) tightens the MBH pair and brings them to central positions \citep{Chandrasekhar1943, Binney1987, Antonini2012_DF}. 
If the DF timescale is less than the Hubble time, the MBH pair may become a gravitationally bound binary. 
However, in certain cases, such as minor mergers in sufficiently massive galaxies or when the secondary MBH undergoes complete tidal stripping faster than the host galaxy’s dynamical timescale, the DF timescale can be so long that orbital decay stalls at $\sim100$\,pc \citep{Dosopoulou2017, Kelley2017}. 

As the binary becomes `harder', stars in the so-called `loss-cone' of low-angular-momentum stellar orbits are the primary scatterers.
In gas-poor mergers, the loss-cone will be depleted if it is only replenished via two-body relaxation, and the binary shrinkage may stall --- once known as the `final parsec problem' \citep{Milosavljevic2003}. 
However, a number of studies have shown that most galaxies are sufficiently tri-axial (i.e., non-spherically symmetric) that the loss-cone can be efficiently replenished \citep{Yu2002_bbh_evolution, Khan2013, Vasiliev2015, Gualandris2017}.  
Eventually, gravitational wave (GW) radiation drives MBH binaries to coalescence, making them the primary sources for Pulsar Timing Array \citep{Burke-Spolaor2019} and the upcoming Laser Interferometer Space Antenna \citep{Amaro-Seoane2023}. In some cases, the GWs carry enough linear momentum to impart a substantial kick to the newly merged MBH, creating a recoiling MBH at off-nuclear positions (e.g., \citealt{Blecha2016}). 

Whether or not MBH binaries can be brought close enough to the GW regime from loss-cone refilling, their lifetimes are long ($\sim{\rm Gyr}$), and a third MBH can enter the system in a subsequent galaxy merger. In such cases, close triple interactions will eject the least massive black hole \citep{Hoffman2007, Bonetti2018_paper3, Ryu2018}, giving it a `slingshot' kick and producing an offset wandering MBH.

The demographics of offset MBHs provides key information on \ad{the formation of MBH mergers \citep{Di-Matteo2023}, ultra-compact dwarf galaxies \citep{Pfeffer2014, Mayes2024}, and} the MBH-host co-evolution paradigm \citep{Volonteri2008}. \ad{Direct evidence of offset MBHs comes from dynamical mass measurements, which requires high angular resolution observations to probe regions within the MBH's sphere of influence $r_{\rm infl}\equiv GM_{\rm BH}/\sigma_\ast^2 \approx 1.5\,{\rm pc}\,(M_{\rm BH}/10^6\,M_\odot)^{0.55}$, where $\sigma_\ast$ is the velocity dispersion of the surrounding stars. 
Using this method, eight MBHs have been detected at the centers of stripped nuclei residing in the halos of their host galaxies \citep{Seth2014, Ahn2017, Ahn2018, Afanasiev2018, Pechetti2022, Voggel2022, Haberle2024}, including one with a sub-kpc offset from its galaxy center \citep{Voggel2022}.
However, this technique is constrained to local galaxies within tens of megaparsecs.}

Another method for detecting offset MBHs have relied on searches for dual/binary active galactic nuclei (AGN) and offset AGN \citep{Comerford2015, DeRosa2019, Hogg2021, Ward2021}. These approaches face substantial selection effects \citep{Van_Wassenhove2012, Blecha2016, Chen2023_dual_offset_agn}, as AGN only probe MBHs that are actively accreting. 
In contrast, tidal disruption events (TDEs) are produced when a star wanders close enough to a MBH to be disrupted, and they occur across all types of galaxies \citep{Sazonov2021, Yao2023, Somalwar2023_vlassI, Masterson2024}. 
Therefore, off-nuclear TDEs offer a unique pathway to probe MBHs irrespective of the state of merger-driven accretion \citep{Ricarte2021_wanderers_tde}. 

\ad{To date, only two\footnote{\ad{While the X-ray outbursts from the IMBH candidate ESO 243-49 HLX-1 \citep{Farrell2009} have been attributed to tidal stripping of a star in an eccentric orbit \citep{Lasota2011}, this interpretation remains debated (e.g., \citealt{Soria2017}) and we do not discuss it here as an off-nuclear TDE.}} off-nuclear TDEs have been identified, both discovered in the X-ray band: 3XMM\,J215022.4–055108 (hereafter 3XMM\,J2150; \citealt{Lin2018_3XMMJ2150, Lin2020}) and EP240222a \citep{Jin2025}. 
In pre-flare optical images, both events are spatially associated with resolved sources, likely stripped satellite dwarf galaxies located in the outskirts of larger parent galaxies.}
Systematically identifying a sample of offset TDEs opens the avenue to addressing key questions in astrophysics that AGN-based studies cannot fully answer, such as mapping the frequency of MBH pairs in diverse galactic environments, assessing the role of galaxy mergers in seeding off-nuclear wandering MBHs \citep{Ricarte2021_wanderers_tde}, and constraining GW kick velocities across the galaxy population \citep{Stone2011}.

The dominant mechanism for generating TDEs is thought to be two-body relaxation \citep{Magorrian1999, Merritt2013, Stone2016}. If the tidal radius ($R_{\rm T}\propto M_{\rm BH}^{1/3}$) is within the BH's event horizon radius ($\propto M_{\rm BH}$), the star will be swallowed without producing an observable flare \citep{Hills1975}. This constraint sets an upper black hole mass limit of $\sim 10^8\,M_\odot$ (for solar-type stars) for detectable TDEs. 
The observed nuclear TDE rate in a typical $10^{10}\,M_\odot$ galaxy is $\sim 3\times10^{-5}\,{\rm galaxy^{-1}\,yr^{-1}}$ \citep{Yao2023}, which aligns well with recent theoretical calculations \citep{Pfister2020, Polkas2024, Teboul2024, Hannah2024_code}. 

It has been known that the observed nuclear TDE rates are elevated in `E+A', compact `green valley' and `post-starburst' galaxies \citep{Arcavi2014, French2016, Law-Smith2017, Graur2018, Hammerstein2021, Sazonov2021, Yao2023}. Such galaxies might be formed by galaxy mergers \citep{Zabludoff1996, Yang2008, Li2023_psb_galaxies}. 
During the DF-dominated phase of a merger, nuclear starburst can enhance the stellar density around the secondary MBH or place stars onto preferentially radial orbits, increasing the TDE rate for $\sim100$\,Myr \citep{Stone2016, Stone2018, Pfister2019_tde_rate, Pfister2021}. 
Once an MBH binary forms, eccentric Kozai-Lidov (EKL; \citealt{Naoz2016}) mechanism and chaotic three-body scatterings can drive stars onto highly eccentric orbits, further boosting the TDE rate around the secondary MBH for 0.1--1\,Myr \citep{Chen2009_tde_rate, Chen2011_tde_rate, Li2015_EKL, Mockler2023, Melchor2024}. A GW-recoiling MBH may trigger the formation of an eccentric disk \citep{Akiba2021}, giving rise to a brief burst of TDE rate \citep{Stone2011, Stone2012, Madigan2018}. 

\subsection{AT2024tvd}

In this Letter we present AT2024tvd, the first off-nuclear TDE selected with optical time-domain surveys. 
AT2024tvd was initially reported to the Transient Name Server as ZTF22aaigqsr \citep{Sollerman2024} based on a detection on 2024 August 25 at $g_{\rm ZTF} = 19.68\pm0.23$\,mag as part of the Zwicky Transient Facility (ZTF; \citealt{Graham2019,Bellm2019b,Dekany2020}) high-cadence partnership survey \citep{Bellm2019-sched} with the 48-inch Samuel Oschin Schmidt telescope at Palomar Observatory (P48). On 2024 August 30 and 2024 October 1, this object passed the thresholds for the ZTF Bright Transient Survey (BTS; \citealt{Fremling2020, Perley2020, Rehemtulla2024}) and Census of the Local Universe (CLU; \citealt{De2020_CLU}) experiments, respectively. 

\begin{figure*}[htbp!]
    \centering
    \includegraphics[width=0.9\textwidth]{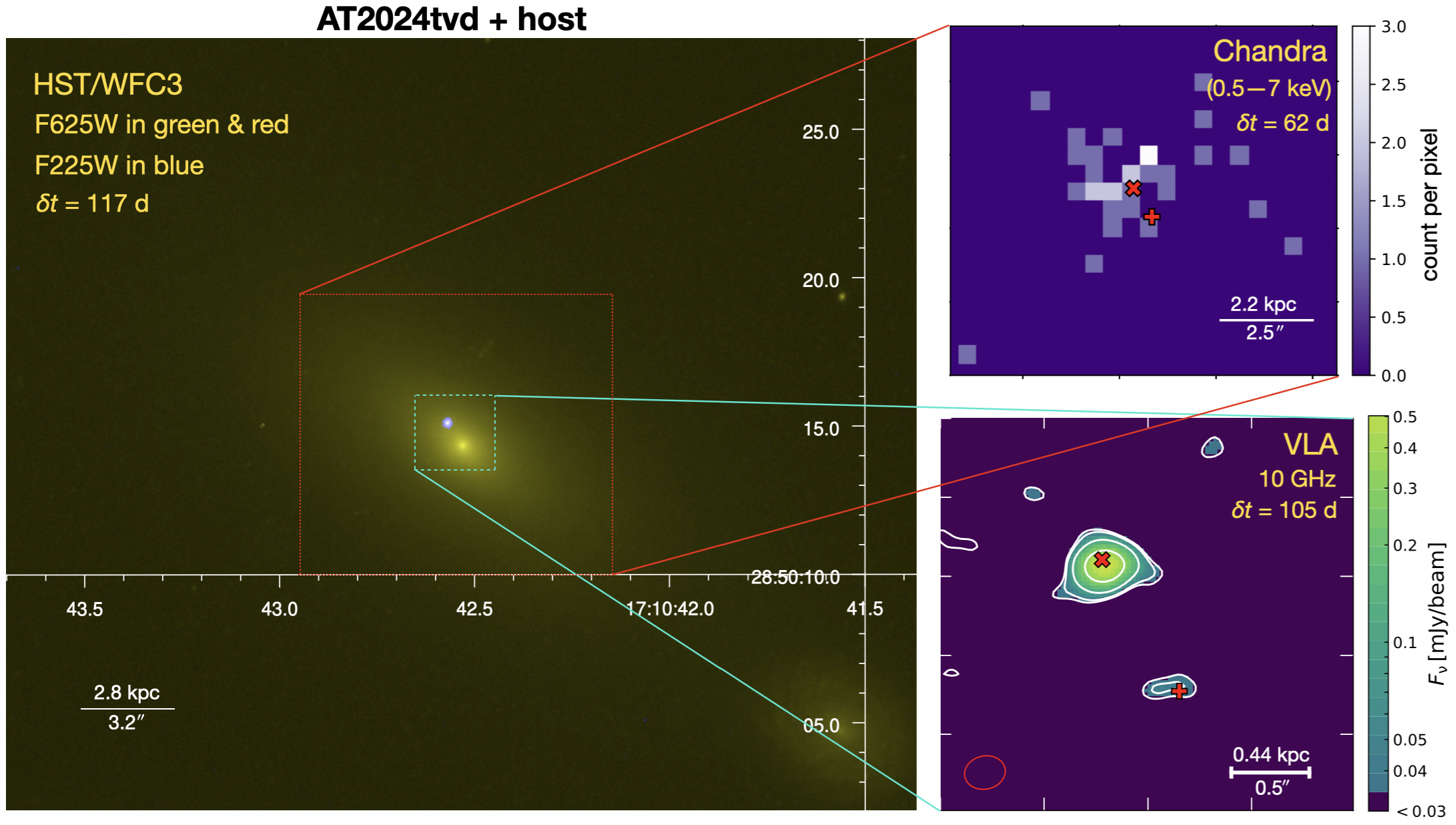}
    \caption{Multi-wavelength images in the field of AT2024tvd. 
    \textit{Left}: The HST image, showing that the transient is offset from the host galaxy nucleus. 
    \textit{Upper Right}: The Chandra (0.5--7\,keV) image. We mark the locations of AT2024tvd (red cross) and host nucleus (red plus) determined by the HST observations. 
    Chandra astrometry indicates that the X-ray source is most likely associated with AT2024tvd (see Section~\ref{subsubsec:cxo_astrometry}).
    \textit{Lower Right}:
    VLA 10\,GHz image. Radio emission is detected from both AT2024tvd and the host nucleus (see Section~\ref{subsec:vla}).
    The white lines mark the 3, 4, 10, and 30 sigma contours of the radio image.
    The red circle is the VLA clean beam.
    \label{fig:image} }
\end{figure*}

On 2024 October 14, AT2024tvd was classified as a TDE by \citet{Faris2024} based on ``broad H and \ion{He}{II}\footnote{Although \citet{Faris2024} reported broad helium lines, our analysis does not provide conclusive evidence for the presence of helium (see Section~\ref{subsec:opt-spec}).} in the spectrum, central location in its host galaxy, and the long-lasting UV detection''. Following this classification report, we noticed that AT2024tvd did not pass the nuclear TDE selection filter developed by the ZTF team. This filter, originally described in \citet{vanVelzen2019} and migrated from the \texttt{AMPEL} broker \citep{Nordin2019} to the \texttt{Kowalski}\footnote{\url{https://github.com/skyportal/kowalski}} broker in September 2023, requires at least one detection alert with a distance between the location of the nearest source in the reference frame and the location of the transient that was smaller than $0.6^{\prime\prime}$. 
An assessment of the ZTF position of AT2024tvd reveals that none of its detection alerts survived the above criterion, and that it is $\sim1^{\prime\prime}$ offset from the nucleus of the host galaxy (\citealt{Yao2025_24tvd_discovery_astronote}, Section~\ref{subsec:opt_phot}). 
Follow-up observations reported here confirm the offset location and the TDE nature (see Figure~\ref{fig:image} and details in Section~\ref{sec:obs}). 

This Letter focuses on data obtained for this transient prior to 2025 January \ad{20}. 
We adopt a standard $\Lambda$CDM cosmology with matter density $\Omega_{\rm M} = 0.3$, dark energy density $\Omega_{\Lambda}=0.7$, and the Hubble constant $H_0=70\,{\rm km\,s^{-1}\,Mpc^{-1}}$.
Times are given in UTC. 
Coordinates are reported in the ICRS frame and J2000 equinox. 
Magnitudes are given in the AB system.
We use the extinction law from \citet{Cardelli1989} with $R_V=3.1$, and adopt a Galactic extinction of $E_{B-V, \rm MW}=0.043$\,mag \citep{Schlafly2011}.
Unless otherwise noted, uncertainties are reported at 1$\sigma$ Gaussian equivalent, and upper limits are reported at 3$\sigma$.

\section{Archival Analysis of Host Galaxy} \label{sec:host}

\subsection{SDSS Spectrum} \label{subsec:sdss_spec}

A host galaxy spectrum was obtained by the Sloan Digital Sky Survey (SDSS, \citealt{Gunn2006}) on 2002 May 16. 
The SDSS DR16 pipeline measures a host redshift of $z=0.04494\pm0.00001$ and a stellar velocity dispersion of $\sigma_\ast = 192.74\pm5.11\,{\rm km\,s^{-1}}$ \citep{Ahumada2020}. 
Using the \ad{\citet{Greene2020}} $M_{\rm BH}$--$\sigma_\ast$ scaling relation \ad{for early-type galaxies}, this implies a central black-hole mass of \ad{${\rm log} (M_{\rm BH}/M_\odot) = 8.37\pm0.08\,({\rm  stat})\pm0.43\,({\rm sys})$}.

\subsection{Host SED Model} \label{subsec:host_sed}
We constructed the pre-TDE photometric spectral energy distribution (SED) from the Galaxy Evolution Explorer (GALEX; \citealp{Martin2005}), SDSS, the Two Micron All-Sky Survey (2MASS; \citealt{Skrutskie2006}) extended source catalog, and the AllWISE catalog \citep{Cutri2014}. 
The GALEX photometry was obtained using the \texttt{gPhoton} package \citep{Million2016} using a radius of 10$^{\prime\prime}$, which is the Kron radius of the galaxy as measured by the Panoramic Survey Telescope and Rapid Response System Data Release 1 (PS1; \citealt{Chambers2016a}) survey.

\begin{figure}[htbp!]
    \centering
    \includegraphics[width=\columnwidth]{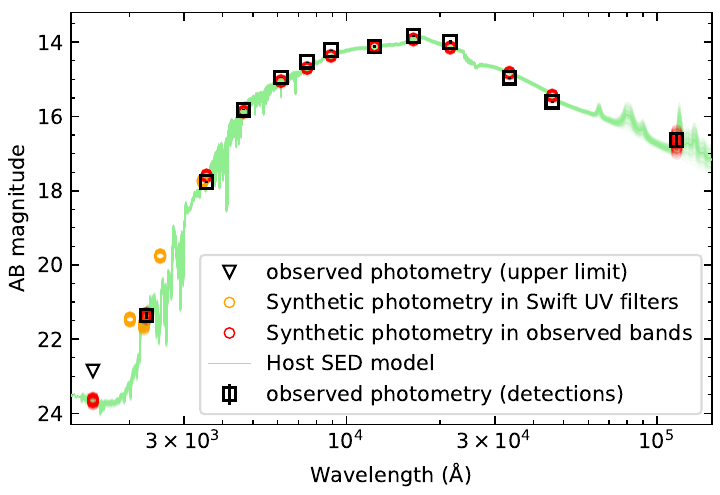}
    \caption{Host galaxy SED of AT2024tvd. 
    \label{fig:host_sed}}
\end{figure}

We modeled the host SED with the flexible stellar population synthesis (\texttt{FSPS}) code \citep{Conroy2009}, adopting an exponentially declining star-formation history (SFH) characterized by the $e$-folding timescale $\tau_{\rm SFH}$, the \citet{Calzetti2000} dust model, and the \citet{Chabrier2003} initial mass function. 
The \texttt{Prospector} package \citep{Johnson2021} was utilized to run a Markov Chain Monte Carlo sampler \citep{Foreman-Mackey2013}.

From the marginalized posterior probability functions we obtain
the total galaxy stellar mass log$(M_{\rm gal}/M_\odot) = 10.93\pm0.02$, 
the metallicity, ${\rm log}Z = -0.41\pm0.03$, 
$\tau_{\rm SFH} = 0.09_{-0.06}^{+0.23}$\,Gyr, 
the population age, $t_{\rm age} = 13.2_{-0.6}^{+0.4}$\,Gyr, 
and negligible host reddening ($E_{B-V, \rm host} = 0.009\pm0.004$\,mag).
In Figure~\ref{fig:host_sed}, the green lines are samples from the posterior distribution of host galaxy SED models.

Using the $M_{\rm gal}$--$M_{\rm BH}$ scaling relation presented in \citet{Greene2020} \ad{for early-type galaxies}, we estimate the black hole mass in the host nucleus to be \ad{${\rm log}(M_{\rm BH}/M_\odot)=8.49\pm0.66$}. 
This is consistent with the black hole mass derived using the $M_{\rm BH}$--$\sigma_\ast$ relation (Section~\ref{subsec:sdss_spec}).

\subsection{Host Type and Centroid} \label{subsec:host_center}
\begin{figure}[htbp!]
    \centering
    \includegraphics[width=0.9\columnwidth]{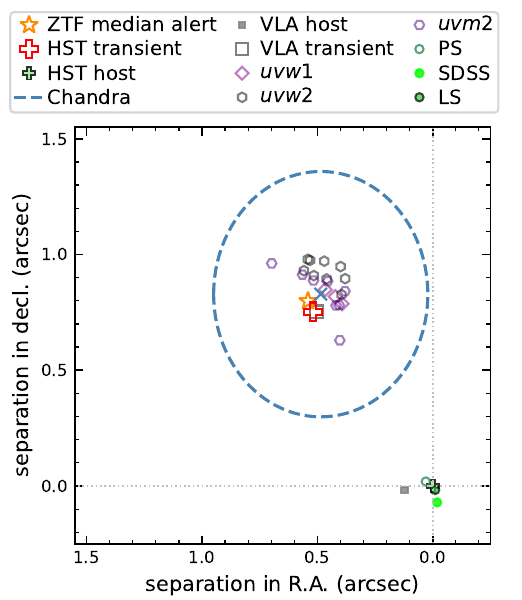}
    \caption{Locations of the host nucleus  and the transient as measured by ZTF alerts (Section~\ref{subsec:opt_phot}), Swift/UVOT (Section~\ref{subsec:uvot}), HST (Section~\ref{subsubsec:hst_phot}), Chandra (blue cross; Section~\ref{subsubsec:cxo_astrometry}), and VLA (Section~\ref{subsec:vla}).
    The 1-$\sigma$ astrometric uncertainty of Chandra is shown by the dashed ellipse. 
    The origin of (0, 0) marks the apparent center of the host galaxy measured by HST. 
    \label{fig:offset}}
\end{figure}

The host of AT2024tvd appears to be a lenticular (S0-type) galaxy with a well-defined centroid. 
Using the SDSS image, \citet{Simard2011} determined that its morphology can be decomposed into a disk component and a bulge component with 
a semimajor effective radius of $R_{\rm e} = (3.29\pm0.02$)\,kpc. 
The galaxy nucleus coordinates given by the DESI Legacy Imaging Survey (LS; \citealt{Dey2019}), PS1, and SDSS are consistent with each other (within a distance of 0.1$^{\prime\prime}$; see the circles in Figure~\ref{fig:offset}). 
The LS host centroid position is
${\rm R.A.}=17^{\rm h}10^{\rm m}42.532^{\rm s}$, 
${\rm decl.}=+28^{\circ}50^{\prime}14.294^{\prime\prime}$. 

\subsection{Galaxy profile modeling} \label{subsec:archival_image_analysis}

In order to search for evidence of previous merger activity or the presence of an additional nuclear star cluster (NSC) offset from the host galaxy center, we undertook modeling of the coadded $g$, $r$ and $z$ band imaging available from the LS DR10 and the coadded $g$, $r$, $i$, $z$ and $y$ PS1 images using the \texttt{Scarlet} multi-band scene modeling software\footnote{https://pmelchior.github.io/scarlet/} \citep{Melchior2018}. Note that the typical size of NSCs is 5\,pc \citep{Neumayer2020}, which corresponds to 6\,mas at $z=0.045$. If an off-nuclear NSC exists and is above the sensitivity limit, we only expect it to be detected as a point source. 

The LS DR10 has a $0.262^{\prime\prime}$ pixel scale and depths of $g\approx 24.7$, $r\approx 23.9$, $z\approx 23.0$\,mag; the PS1 has a $0.258^{\prime\prime}$ pixel scale and depths of $g\approx 23.3$, $r\approx 23.2$, $i\approx 23.1$, $z\approx 22.3$, and $y\approx 21.3$\,mag. 
For the LS DR10 models, we provided \texttt{Scarlet} with the point source function (PSF) model images provided by the data release.  For the PS1 imaging, we reconstructed the PSF image using the best-fit PSF parameters as described in \citet{Magnier2020} and published in the StackObjectAttributes table in the online PS1 catalog \citep{Flewelling2020}. 
In each case we ran Source Extractor \citep{Bertin1996} to identify all sources detected over a $5\sigma$ threshold. We required that the galaxy models be monotonically decreasing --- but not radially symmetric --- and that they have the same morphology in each band (such that the SED does not vary in different regions of the galaxy). This enables us to avoid any assumptions about the galaxy following an analytical galaxy profile. \texttt{Scarlet} was run to convergence 
to fit the multi-band SEDs and galaxy morphologies for the sources in the scene. 

The best fit model, corresponding observations, and residuals are shown in Appendix~\ref{sec:figures} (Figure \ref{fig:Scarletmodel}). Neither the LS residuals nor the PS1 residuals show any evidence of tidal tails or asymmetric structures in the galaxy to suggest pre-merger activity. In the LS $g$-band image, we estimate the limiting magnitude in the Galactic bulge by determining the pixel variance of the residuals in a 30$\times$30 pixel cutout centered on the TDE position (see Section~\ref{subsubsec:hst_phot}). We determine a limiting magnitude of $g\approx 23.19$\,mag, which implies that no NSC exists at the TDE position with an absolute $g$-band magnitude brighter than $-13.30$\,mag. 
This limit excludes only the most luminous NSCs \citep{Neumayer2020}. \ad{For comparison, the satellite dwarf galaxy associated with EP240222a was detected in LS imaging with $M_g=-11.81$\,mag and has a stellar mass of $\sim10^{7.0}\,M_\odot$ \citep{Jin2025}. Assuming a similar $g$-band mass-to-light ratio, the mass of the possible star cluster associated with AT2024tvd can be constrained to be $< 10^{7.6}\,M_\odot$}.

\subsection{eROSITA X-ray Upper Limit} \label{subsec:eROSITA}

The position of AT2024tvd was observed by the eROSITA telescope \citep{Predehl2021} on board the \textit{Spektrum-Roentgen-Gamma} (\textit{SRG}) satellite \citep{Sunyaev2021}. 
The position was observed on four epochs from 2020 to 2022, each separated by approximately 6\,months, with the first observation held on 2020 March 13. 
\ad{No X-ray photons were detected by eROSITA within a $15^{\prime\prime}$ radius centered on the position of AT2024tvd. To convert the upper limit on the count rate to a flux, we assume an absorbed power-law spectrum with Galactic column density fixed at $N_{\rm H}=4.43\times 10^{20}\,{\rm cm^{-2}}$ and a photon index of $\Gamma=3$ (see Section~\ref{subsec:xrt}). }
The 0.2--2.3\,keV upper limit at the 90\% confidence is $\sim 3.5\times 10^{-14}\,{\rm erg\,s^{-1}\,cm^{-2}}$ in individual scans, and $\sim 2.0\times 10^{-14}\,{\rm erg\,s^{-1}\,cm^{-2}}$ in the combined data of all four observations. 
The latter upper limit corresponds to $L_{\rm X}<9.6\times 10^{40}\,$erg\,s$^{-1}$. 

\section{New Observations and Analysis} \label{sec:obs}

\subsection{ZTF and ATLAS}
\label{subsec:opt_phot}

We obtained ZTF \citep{Masci2019, Masci2023} and ATLAS \citep{Tonry2018, Smith2020, Shingles2021} forced photometry using the median position of ZTF alerts generated for AT2024tvd before November 2024: 
${\rm R.A.}=17^{\rm h}10^{\rm m}42.574^{\rm s}$, 
${\rm decl.}=+28^{\circ}50^{\prime}15.110^{\prime\prime}$. 
Baseline correction was performed following the procedures outlined in \citet{Yao2019}. The Galactic extinction-corrected optical light curves are shown in Figure~\ref{fig:opt_lc} and presented in Appendix~\ref{sec:table} (Table~\ref{tab:phot}).

\begin{figure}[htbp!]
\centering
    \includegraphics[width=\columnwidth]{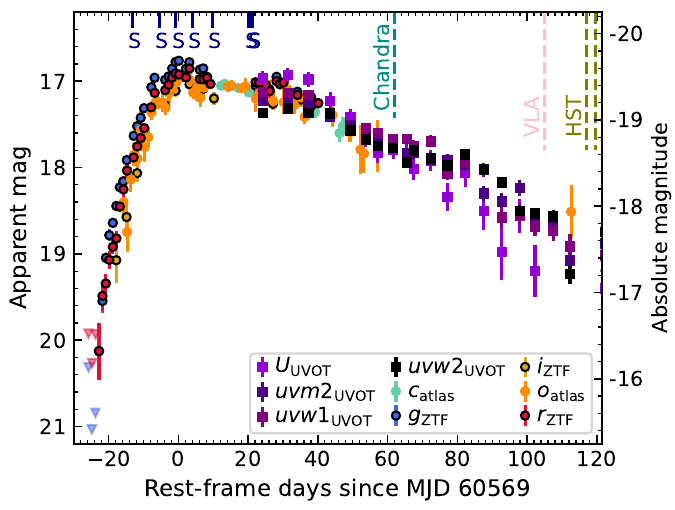}
    \caption{The optical and UV light curve of AT2024tvd, with epochs of optical spectroscopy marked with letter `S'. The Chandra, VLA, and HST observing epochs are also indicated.
    \label{fig:opt_lc}}
\end{figure}

The $g_{\rm ZTF}$-band light curve reaches a maximum on 2024 September 16 (MJD 60569). Hereafter we use $\delta t$ to denote rest-frame days relative to this epoch. 

The ZTF alert photometry median position reported above is 0.98$^{\prime\prime}$ away from the LS host galaxy centroid (Section~\ref{subsec:host_center}). However, this alone does not definitively imply that AT2024tvd is an off-nuclear transient, because in rare cases, the alert astrometry can be off by as large as $\sim1^{\prime\prime}$ (see ZTF19acymzwg analyzed in \citealt{Brightman2021} as an example).
To robustly assess the positional offset, we undertook modeling of the multi-epoch ZTF imaging using the scene modeling code \texttt{Scarlet2}\footnote{https://github.com/pmelchior/scarlet2}, which can model varying point sources against a static background \citep{Sampson2024, Ward2024}. 
The modeling procedure, detailed in Appendix~\ref{sec:ztf_astrometry}, yields an offset of $0.95\pm0.42^{\prime\prime}$ ($3\sigma$ uncertainty) between AT2024tvd and the host galaxy nucleus. This confirmed the offset location at a significance of $6.8\sigma$, and motivated higher-resolution imaging observations. 

\subsection{HST} 
We observed the field of AT2024tvd under a DDT program (PI Y. Yao) using the \textit{Hubble Space Telescope} (HST).

\subsubsection{HST Imaging} \label{subsubsec:hst_phot}
HST imaging observations were conducted on 2025 January 16 ($\delta t = 117$\,d) using the Wide Field Camera 3 (WFC3) with two bands: F225W and F625W. These data can be found in MAST: \dataset[10.17909/f0s7-mn70]{http://dx.doi.org/10.17909/f0s7-mn70}.

An HST color image is presented in the left panel of Figure~\ref{fig:image}. In the F625W image, the transient is clearly offset from the host galaxy nucleus, while in the F225W image, only the transient is detected. This explains why the transient appears white, whereas the host galaxy appears yellow. 

In order to determine the transient and host nucleus positions, we modeled the galaxy profile and the TDE in the HST F625W image. This also allows us to search for extended emission around the TDE and any associated tidal structures. We first constructed a PSF model by identifying the three stars in the image that were uncontaminated by any galaxy profiles and taking a median stack of cutouts of the stars. As the stars were at low signal-to-noise ratio (S/N) and only three were available, some noise is introduced by the PSF model. We again used \texttt{Scarlet} as described in Section \ref{subsec:archival_image_analysis} to model the large host galaxy as (i) a Spergel (pseudo-S\'{e}rsic) profile \citep{Spergel2010} and (ii) a non-parametric, monotonically decreasing profile. In addition, we modeled the TDE as a single point source. The models, observations, and residuals are shown in Appendix~\ref{sec:figures} (Figure \ref{fig:HSTScarletmodel}). No NSC centered on the TDE, or tidal structures, are visible. 

We find the best-fit host galaxy center and its 3$\sigma$ uncertainty from the parametric fit to be ${\rm R.A.}=17^{\rm h}10^{\rm m}42.53271
\pm0.00011^{\rm s}$, 
${\rm decl.}=+28^{\circ}50^{\prime}14.311397\pm0.000094^{\prime\prime}$;
and the best-fit position of AT2024tvd in both the parametric and non-parametric fits to be
${\rm R.A.}=17^{\rm h}10^{\rm m}42.5722\pm0.0065^{\rm s}$, ${\rm decl.}=+28^{\circ}50^{\prime}15.0635\pm0.0011^{\prime\prime}$. 
The HST positions are marked by pluses in Figure~\ref{fig:offset}.
This corresponds to an offset of $0.91380\pm0.00043^{\prime\prime}$ (3$\sigma$ uncertainty). We note that the statistical uncertainty reported for the host galaxy center does not encompass uncertainties introduced by the fact that the parametric profile does not fully describe the data. 
The PSF full-width at half maximum (FWHM) of this image is $0.068^{\prime\prime}$.
To obtain a more conservative estimate of the transient--host offset uncertainty, we take the positional uncertainty of the transient and the host nucleus to be 10\% of the PSF FWHM. The offset is found to be $0.914\pm0.010^{\prime\prime}$.

We also modeled the galaxy light profile using \texttt{GALFIT} \citep{Peng2010} to obtain galaxy components similar to previous TDE host modeling studies \citep[e.g.,][]{Law-Smith2017, Hammerstein2023, Hammerstein2023_IFU}. In each case, we model the TDE as a single point source using the PSF created from the stars in the image. For completeness we also model the nearby companion galaxy, which is relatively well-fit by an exponential disk model with S\'ersic index $n\approx 1$. This is expected given the appearance of spiral structure, presumably associated with a disk, in the HST F625W image. To model the transient host, we first fit a single S\'ersic model, which yields $n=5.259 \pm 0.005$. This is similar to single S\'ersic fits to archival SDSS imaging for other TDE hosts \citep{Law-Smith2017, Hammerstein2023}, which implies a more centrally-concentrated morphology. The shape of the host forces the entire single S\'ersic component to appear more elongated. In order to avoid this, we fit a double S\'ersic model which captures the central shape of the host more accurately. This fit yields a bulge S\'ersic index of $n_{\rm bulge} = 2.953 \pm 0.010$ and a disk S\'ersic index of $n_{\rm disk} = 1.172 \pm 0.003$. We do not find significant evidence for extended emission above the noise of the PSF model at the location of the transient. The models, observations, and residuals are shown in the Appendix~\ref{sec:figures} (Figure \ref{fig:HSTgalfitmodel}).

\subsubsection{HST Spectroscopy} \label{subsubsec:hst_spec}
\ad{HST UV spectroscopic observations were conducted on 2025 January 19 ($\delta t = 120$\,d) using the Space Telescope Imaging Spectrograph (STIS) with the NUV and FUV MAMA detectors. 
The spectra were obtained through a $52^{\prime\prime} \times 0.2^{\prime\prime}$ aperture. A nearby offset star was used for source acquisition. 
We used the G140L and G230L gratings to cover the spectral ranges 1150--1730\AA\ and 1570--3180\AA, respectively.
The observations spanned three orbits, with total exposure times of 4629\,s in the FUV and 2624\,s in the NUV.  
These data can be found in MAST: \dataset[10.17909/pnnr-xm74]{http://dx.doi.org/10.17909/pnnr-xm74}.
}

\begin{figure*}[htbp!]
    \centering
    \includegraphics[width=\textwidth]{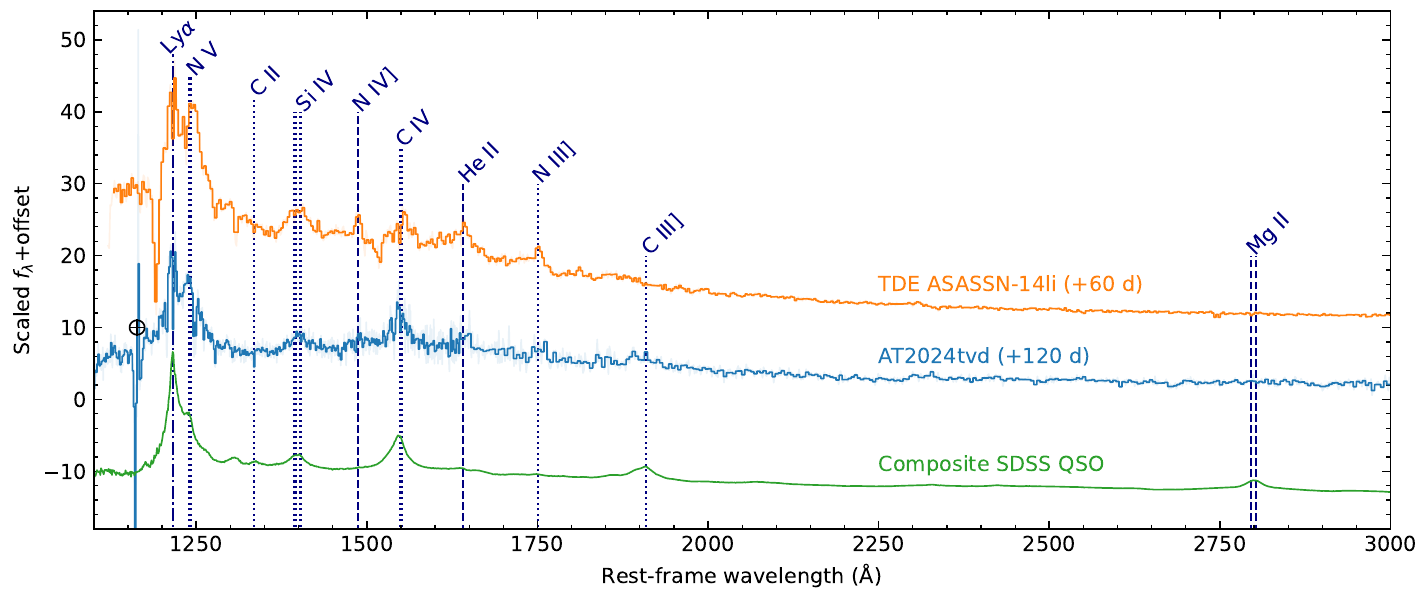}
    \caption{\ad{The HST UV spectrum of AT2024tvd (shifted to rest-frame wavelength using the host redshift of $z=0.04494$), compared with the HST spectrum of ASASSN-14li \citep{Cenko2016} and the composite QSO spectrum from SDSS \citep{VandenBerk2001}. The spectral features of AT2024tvd closely resemble those seen in the TDE ASASSN-14li.}\label{fig:hst_spec}}
\end{figure*}

\ad{We combined the one-dimensional (1D) spectra using inverse-variance weighting, and the resulting spectrum is shown in Figure~\ref{fig:hst_spec}. We detected narrow absorption lines at both $z=0$ and the host redshift, as well as broad emission features at the host redshift. 
The broad features, including Ly$\alpha$,
\ion{N}{V} $\lambda\lambda$1239, 1243,
\ion{Si}{IV} $\lambda\lambda$1394, 1403, 
\ion{C}{IV} $\lambda\lambda$1548, 1551, 
\ion{He}{II} $\lambda1642$,
and \ion{N}{III}] $\lambda$1750, closely resemble those observed in the ``Rosetta Stone'' TDE ASASSN-14li \citep{Cenko2016} and must originate from AT2024tvd itself.
Their presence provides strong support for the TDE nature of AT2024tvd, and confirms its association with the host galaxy, ruling out the possibility of it being a foreground or background transient. }

\subsection{Swift/XRT} \label{subsec:xrt}
AT2024tvd was observed by the X-ray Telescope (XRT; \citealt{Burrows2005}) and the Ultra-Violet/Optical Telescope (UVOT; \citealt{Roming2005}) on board the \textit{Neil Gehrels Swift Observatory} under a series of time-of-opportunity requests starting on 2024 October 10. We process the Swift data using \texttt{HEASoft} version 6-33.2 and CALDB version 1.0.2. 

All XRT observations were obtained in the photon-counting mode. 
First, we ran \texttt{ximage} to determine the position of AT2024tvd in each observation. 
To calculate the background-subtracted count rates, we filtered the cleaned event files using a source region with $r_{\rm src} = 40^{\prime\prime}$, and eight background regions with $r_{\rm bkg} = 30^{\prime\prime}$ evenly spaced at $100^{\prime\prime}$ from AT2024tvd.

For each observation, we generated a spectrum with \texttt{xselect}, and applied optimal binning with \texttt{ftgrouppha} \citep{Kaastra2016} while ensuring at least one count per bin. 
For \ad{adjacent} observations \ad{with low counts}, we stacked the data together to increase the S/N. We verified that pile-up is not present in the observation with the highest count rate (obsID 16860016).

\begin{figure}[htbp!]
    \centering
    \includegraphics[width=\columnwidth]{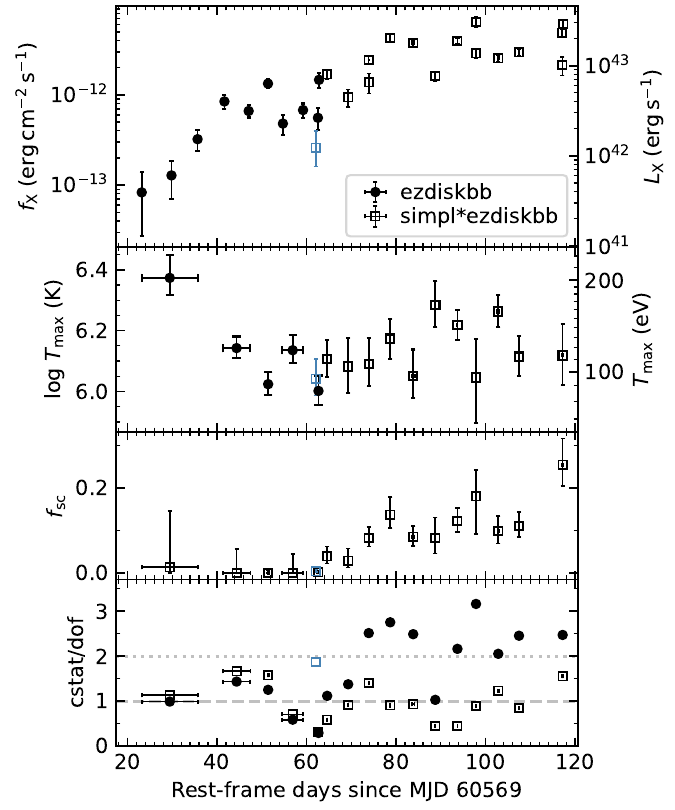}
    \caption{X-ray light curve, spectral model parameters, and fit statistics. The Swift/XRT results are shown in black, and the Chandra result is shown in blue.  \label{fig:xrt_lc}}
\end{figure}

\ad{We determined the appropriate energy range for spectral fitting by requiring that the net count rate exceeded the background count rate}. We then modeled the spectra with an absorbed multi-temperature disk (i.e., \texttt{tbabs*zashift*ezdiskbb} in \texttt{xspec}).
The Galactic column density $N_{\rm H}$ was fixed at $4.43\times 10^{20}\,{\rm cm^{-2}}$ \citep{HI4PI2016}. 
\ad{The \texttt{ezdiskbb} model assumes zero torque at the inner disk edge and has two parameters: the maximum temperature in the disk ($T_{\rm max}$) and a normalization term \citep{Zimmerman2005}.}
The data were fitted using $W$-statistics via \texttt{cstat} \citep{Cash1979}. 
At $\delta t>70$\,d, the fit statistics (\texttt{cstat/dof}) generally exceeded two (see the bottom panel of Figure~\ref{fig:xrt_lc}), and a hard tail emerged that was not captured by the disk model.

\ad{To account for this hard excess, we added a \texttt{simpl} component \citep{Steiner2009}, which assumes that a fraction ($f_{\rm sc}$) of the thermal seed photons are inverse Compton scattered to produce a hard power-law component with a photon index of $\Gamma$. Leaving both parameters free often led to them being unconstrained. Therefore, we fixed $\Gamma$ at 2.5, a typical value found in TDEs with a prominent power-law component \citep{Guolo2024}, and fit only for $f_{\rm sc}$.}

\ad{To assess the statistical improvement between the two models, we computed the Bayesian Information Criteria (BIC), where a lower value of ${\rm BIC}= \texttt{cstat} + k \cdot {\rm ln}(N) $ indicates a preferred model. Here $N$ is the number of spectral bins, and $k$ is the number of free parameters. 
We found that the pure disk model yields slightly lower BIC values in the first five spectra, and much higher BIC values in later observations. 
This can be understood by examining the $f_{\rm sc}$ evolution shown in Figure~\ref{fig:xrt_lc}, which initially remains consistent with zero and increased to 10--20\% at later times.}

\ad{Next, we tested for intrinsic absorption at the host redshift. For each spectrum, we convolved the BIC-selected model with a \texttt{ztbabs} component, and recorded the best-fit BIC values. 
In all cases, the best-fit $N_{\rm H, host}$ was consistent with zero, and the BIC increased relative to the reference model. 
We thus conclude that no significant intrinsic absorption is present.}

\ad{Finally,} for each obsID, we computed the 0.3--10\,keV net count rate to flux conversion factors using the model with \ad{the lower BIC value}.
To generate the XRT light curve, we calculated 0.3--10\,keV net count rates by filtering the cleaned event files using the same source and background regions described above. 
We first binned the light curve by good time intervals (GTIs), with each obsID containing 1–4 GTIs. 
If, within a single obsID, the difference in count rates across consecutive GTIs was less than 2$\sigma$, we further combined the GTIs together. 
The net count rates were then converted to fluxes using the conversion factors.
We uncovered significant variability (i.e., at least a multiplicative factor of two in flux change) on short timescales (from 1.6\,hr to a few days). 

The X-ray light curve and best-fit spectral parameters are shown in Figure~\ref{fig:xrt_lc} and given in Appendix~\ref{sec:table} (Table~\ref{tab:xrt}). 

\subsection{Swift/UVOT}\label{subsec:uvot}
We measured the UVOT photometry using the \texttt{uvotsource} tool. We used a circular source region with $r_{\rm src} = 10^{\prime\prime}$, and corrected for the enclosed energy within the aperture. 
We measured the background using two nearby circular source-free regions with $r_{\rm bkg} = 15^{\prime\prime}$. 

We estimated host galaxy flux in the UVOT filters from the best-fit host SED model (see Section~\ref{subsec:host_sed}), which gives observed magnitudes of $uvw2=21.73$\,mag, $uvm2=21.96$\,mag, $uvw1=19.99$\,mag, and $U = 17.95$\,mag. 
Considering the significant host contribution in $B$ and $V$ bands and the uncertainties in the host SED model, we exclude these two bands from our analysis.
The host-subtracted UVOT photometry is shown in Figure~\ref{fig:opt_lc} and given in Appendix~\ref{sec:table} (Table~\ref{tab:phot}).

UVOT also allows us to estimate the location of the UV emitting region of AT2024tvd. 
To this end, we selected images where the host-subtracted transient flux is brighter than 200\,$\mu$Jy and at least a factor of ten greater than the galaxy SED model prediction. 
This criterion yielded 20 images, including 3 in $uvw1$, 8 in $uvm2$ and 9 in $uvw2$. 

To assess the pointing accuracy of UVOT, we first ran Source Extractor\footnote{We used the \texttt{python} package \texttt{sep\_pjw} \citep{Barbary2016}.} on each image, and then cross-matched the list of detected sources with the Gaia DR3 catalog \citep{Gaia2023_dr3} using a cross-match radius of 1$^{\prime\prime}$.  
The number of matched sources per image ranges from 15 to 86. 
For each image, the UVOT pointing offset and its uncertainty in R.A. and decl. were determined as the median and standard deviation of the differences between the Source Extractor positions and the Gaia positions. 
Typically, the pointing offset is $< 0.1^{\prime\prime}$ in both directions, with an uncertainty of $\sim 0.35^{\prime\prime}$. These pointing offsets were then applied to correct the coordinates of the UV transient. 
The resulting locations of the UV transient across the 20 individual images are shown in Figure~\ref{fig:offset}.
We conclude that the UVOT location is consistent with the HST location of AT2024tvd, and is offset from the host centroid. 

\subsection{Optical and UV Photometric Analysis} \label{subsec:bbfit}

\begin{figure}[htbp!]
    \centering
    \includegraphics[width=\columnwidth]{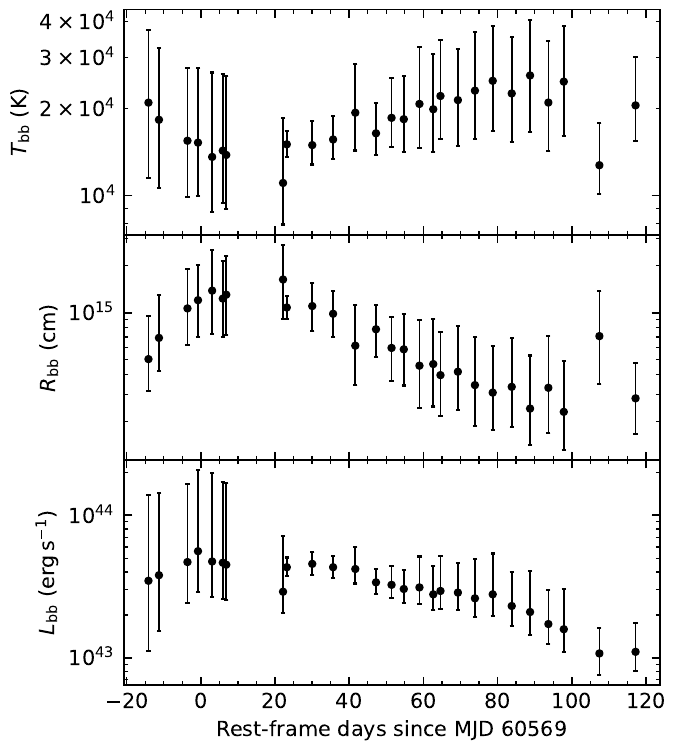}
    \caption{Evolution of the UV/optical blackbody properties of AT2024tvd. \label{fig:bb_pars}}
\end{figure}

We construct a SED for AT2024tvd at each epoch with detections in no less than four filters, and fit a blackbody function following the method adopted by \citet{Yao2020}. 
The SED fits are shown in Appendix~\ref{sec:figures} (Figure~\ref{fig:bbfits}). 
The resulting best-fit blackbody temperature ($T_{\rm bb}$), radius ($R_{\rm bb}$), and luminosity ($L_{\rm bb}$) are shown in Figure~\ref{fig:bb_pars}. 

\subsection{Optical Spectroscopy}
\label{subsec:opt-spec}

We obtained optical spectra with the Spectral Energy Distribution Machine (SEDM; \citealt{Blagorodnova2018, Rigault2019, Kim2022}) on the robotic Palomar 60 inch telescope (P60; \citealt{Cenko2006}), the Double Beam Spectrograph (DBSP; \citealt{Oke1982}) on the Palomar 200-inch Hale telescope (P200), Binospec \citep{Fabricant2019} on the 6.5\,m MMT telescope, and the Kast spectrograph on the Shane 3-m telescope at Lick Observatory \citep{Miller1993}. 
These observations were coordinated using the \textit{fritz.science} instance of \texttt{SkyPortal} \citep{vanderWalt2019, Coughlin2023}.
The SEDM spectroscopic observations were obtained as part of the ZTF BTS.
Epochs of spectroscopic observations are marked with `S' in Figure~\ref{fig:opt_lc}, and observation details are provided in Table~\ref{tab:spectra}. 

\begin{deluxetable}{lrcc}[htbp!]
\tablecaption{Spectroscopic observations of AT2024tvd. 
\label{tab:spectra}} 
\tablehead{\colhead{Start Date} & \colhead{$\delta t$} & \colhead{Tel.+Instr.} & \colhead{Exp.} \\ 
\colhead{(MJD)} & \colhead{(d)} & & \colhead{(s)} } 
\startdata 
60555.2057 & $-13.2$ & P60+SEDM & 2700  \\
60563.2800 & $-5.5$ & P200+DBSP & 1200  \\
60568.1923 & $-0.8$ & P60+SEDM & 2160   \\
60579.1991 & $+9.8$ & P60+SEDM & 2160  \\
60590.1520 & $+20.2$ & Shane+Kast & 1570/1500\tablenotemark{a} \\
60591.0896 & $+21.1$ & MMT+Binospec & 1320  \\
60591.1228 & $+21.2$ & P60+SEDM & 2160  \\
60591.1262 & $+21.2$ & Shane+Kast &  1570/1500\tablenotemark{a}\\
\enddata 
\tablenotetext{a}{Exposure times on blue/red sides of the spectrograph.}
\end{deluxetable}

For the DBSP spectrum, we used the D-55 dichroic filter, the 600/4000 grating on the blue side, the 316/7500 grating on the red side, and a slit width of 2\farcs0. 
The spectrum was reduced using the \texttt{dbsp\_drp} pipeline \citep{Roberson2022}, which is based on \texttt{PypeIt} \citep{pypeit:joss_pub}. 

For the Kast spectra, we used the 300/7500 grating, the 600/4310 grism, and a slit width of 1\farcs5. 
The spectra were reduced using the \texttt{UCSC Spectral Pipeline}\footnote{\url{https://github.com/msiebert1/UCSC\_spectral\_pipeline}} \citep{Siebert20}, a custom data-reduction pipeline based on procedures outlined by \citet{Foley03}, \citet{Silverman2012}, and references therein. The two-dimensional spectra were bias-corrected, flat-field corrected, adjusted for varying gains across different chips and amplifiers, and trimmed. One-dimensional spectra were extracted using the optimal algorithm \citep{Horne1986}. The spectra were wavelength-calibrated using internal comparison-lamp spectra with linear shifts applied by cross-correlating the observed night-sky lines in each spectrum to a master night-sky spectrum. Flux calibration and telluric correction were performed using the high S/N standard Feige 34 observed on the 2\textsuperscript{nd} night. More details of this process are discussed elsewhere \citep{Foley03, Silverman2012, Siebert20}. We then combined the output red and blue spectra by scaling one spectrum to match the flux of the other using the ratio of the mean fluxes of both sides.

For the Binospec spectrum, we used the 270 line grating with a central wavelength of 6500\,\AA, the blocking filter LP3800, and a slit width of $1.0^{\prime\prime}$. 
The data were reduced using \texttt{PypeIt}. In extracting the spectrum, which is blended with the host galaxy, we used the Horne algorithm \citep[the \texttt{optimal} method in \texttt{PypeIt};][]{Horne1986} and forced the FWHM of the trace to be 6 pixels ($\sim$1\farcs4), consistent with the seeing. We note that starting from the 2024B semester, Binospec would randomly suffer a drop in the throughput on the blue side ($\lesssim$5000\,\r{A}) in longslit spectroscopy, possibly associated with a failure in the atmospheric dispersion corrector. We spotted the same issue in this spectrum. Since MMT does not take telluric standard for Binospec every night, we adopted the spectrum of BD+332642 observed on October 4 2024 (fours days before observing AT2024tvd), which suffered a similar drop in throughput, for flux calibration. The resultant 1D spectrum is consistent with spectra obtained with other instruments. Nevertheless, the continuum may still be problematic and we exclude this Binospec spectrum in our SED fitting.

\begin{figure}[htbp!]
\centering
    \includegraphics[width=\columnwidth]{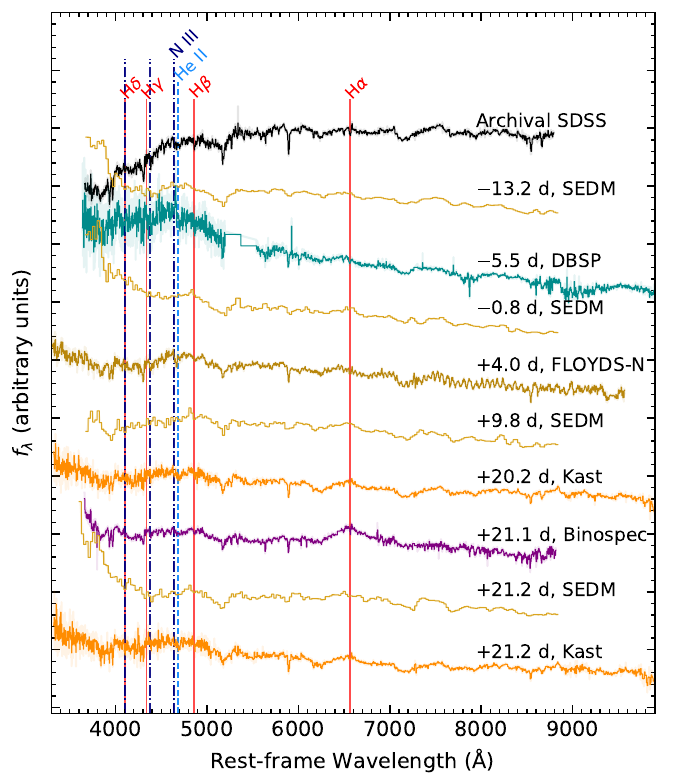}
    \caption{The optical spectra of AT2024tvd. Also shown on the top is an archival SDSS spectrum of the host galaxy (see Section~\ref{subsec:sdss_spec}). \label{fig:opt_spec}}
\end{figure}

The spectral sequence is shown in Figure~\ref{fig:opt_spec}. For comparison, we also show the FLOYDS-N optical spectrum published in \citet{Faris2024}.

\begin{figure}[htbp!]
    \centering
    \includegraphics[width=\columnwidth]{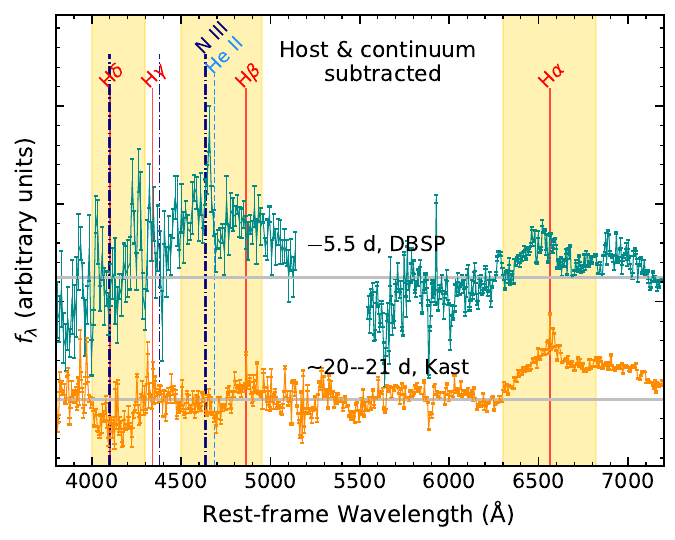}
    \caption{Host- and continuum-subtracted DBSP and Kast spectra. 
    A broad feature around H$\alpha$ is observed in both spectra, while the broad complex around H$\beta$ and \ion{He}{II}+\ion{N}{III} is only observed in the DBSP sepctrum. 
    Wavelength ranges not included in the fitting are marked with the light yellow bands. \label{fig:opt_subtract}}
\end{figure}

It is evident from Figure~\ref{fig:opt_spec} that the transient exhibits a blue continuum and a broad emission line around H$\alpha$. 
The H$\alpha$ line appears to be most prominent in the Binospec spectrum, thanks to the narrow slit used. The $-0.8$\,d SEDM spectrum also clearly displays a broad emission line around H$\beta$. To search for other features, such as the \ion{He}{II} and \ion{N}{III} commonly observed in some TDEs \citep{vanVelzen2021}, we fit the DBSP and the combined Kast spectra in rest-frame 3800--7200\,\AA. This fitting utilized a combination of blackbody emission and host galaxy contributions, following the procedure outlined in \citet{Yao2022}. Wavelength regions where broad lines might appear were excluded from the fit (marked by yellow bands in Figure~\ref{fig:opt_subtract}).
For the blackbody temperature, we allow it to vary within the 68\% confidence intervals of the $T_{\rm bb}$ value shown in Figure~\ref{fig:bb_pars} that is closest to the spectroscopic phase. 
For the host spectrum, we utilized the best-fit host SED model (i.e., green lines shown in Figure~\ref{fig:host_sed}). However, we note that this model represents the global galaxy spectrum and may not accurately reflect the local stellar population at the location of this offset TDE.

The host- and continuum-subtracted spectra are presented in Figure~\ref{fig:opt_subtract}. Both spectra show a broad excess near H$\alpha$. Blueward of $\sim$6000\,\AA, the $\sim$20--21\,d Kast spectrum lacks evident spectral features, whereas the $-5.5$\,d DBSP spectrum reveals a flux excess at 4500--5100\,\AA. This flux excess can be attributed to a combination of a broad H$\beta$ line and a Bowen fluorescence complex, comprising \ion{He}{II} $\lambda4686$ and \ion{N}{III} $\lambda$4640. However, given that the DBSP spectrum has a wavelength gap and is noisy, we are not able to confidently identify the Bowen features. Therefore, we classify AT2024tvd's optical spectroscopic subclass as TDE-H+He or TDE-H, using the nomenclature introduced in \citet{vanVelzen2021}. 

We note that the H$\alpha$ line width in FWHM is $\sim1.4 \times 10^4 \,{\rm km\,s^{-1}}$. As shown by \citet{Charalampopoulos2022}, such a broad width is more commonly observed in TDEs with Bowen lines. 

\subsection{Chandra} \label{subsec:chandra}
AT2024tvd was observed by the Chandra X-ray Observatory under a DDT program (PI Y. Yao) on 2024 November 18 ($\delta t = 62$\,d) for a total of 21.5\,ks (obsID 30620). 
We used the Advanced CCD Imaging Spectrometer (ACIS; \citealt{Garmire2003}), with the aim point on the back illuminated S3 chip. 
We reduced the data using the Chandra Interactive Analysis of Observations (\texttt{CIAO}; \citealt{Fruscione2006}) software package (\ad{v4.17}). 

\subsubsection{\ad{Chandra detection and source profile}}

\ad{An X-ray source is clearly detected around the HST position of the transient. 
To assess whether the source is extended, we first simulated the PSF using the Chandra Ray Tracer (ChaRT; \citealt{Carter2003}). 
We then projected the PSF onto the detector-plane with MARX \citep{Davis2012} via the \texttt{simulate\_psf} command, and created an image of the PSF. 
Finally, we ran \texttt{srcextent} to estimate the source size. 
The observed source size is 0.92$^{\prime\prime}$ (90\% confidence interval: 0.61--1.23$^{\prime\prime}$). 
The PSF observed size is 0.47$^{\prime\prime}$ (90\% confidence interval: 0.40--0.53$^{\prime\prime}$). 
The estimated intrinsic size is 0.79$^{\prime\prime}$ (90\% confidence interval: 0.52--1.07$^{\prime\prime}$). 

The source is not extended at 90\% confidence. As such, we treat the Chandra detection of the transient as a point source throughout the remainder of this manuscript.
}

\subsubsection{Chandra astrometry} \label{subsubsec:cxo_astrometry}

To determine the astrometric shifts of Chandra images, we first ran \texttt{fluximage} to filter 1--7\,keV events, and then ran \texttt{wavdetect} to obtain lists of positions for all sources in the ACIS-S S2 and S3 chips. 
Wavelet scales of 1, 2, and 4 pixels and a significance threshold of $10^{-6}$ were used. 
A total of 23 X-ray sources were detected. 
To assess the Chandra absolute astrometry, we cross matched the X-ray source list with Gaia DR3\footnote{The Gaia object centered on the host nucleus was not included during this cross-matching step.}, using a radius of $2^{\prime\prime}$. 
This left six Chandra/Gaia sources.

Afterwards, we use \texttt{wcs\_match} and \texttt{wcs\_update} to apply a (translation-only) astrometric correction. After the correction, we ran \texttt{wavdetect} again on the corrected X-ray image. 
\ad{Following the method outlined in \citet{Rots2011}, we determine the Chandra systematic uncertainty $\sigma_{\rm sys}$, such that for each matched pair, the normalized angular separation is
\begin{align}
    \rho_{\rm norm} = \frac{\rho}{\sqrt{\sigma_{\alpha, \rm X}\sigma_{\delta, X} + \sigma_{\alpha, \rm opt}\sigma_{\delta, \rm opt}  + \sigma_{\rm sys}^2}}.
\end{align}
Here $\rho$ is the angular separation of X-ray and optical positions; $\sigma_{\alpha, \rm X}$, $\sigma_{\delta, \rm X}$, $\sigma_{\alpha, \rm opt}$, and $\sigma_{\delta, \rm opt}$ are the $1\sigma$ uncertainties in R.A. and decl. for the X-ray and optical positions, respectively. 
Given that $\rho_{\rm norm}$ is expected to follow a Rayleigh distribution with mean of $\sqrt{\pi/2}$, we derive  $\sigma_{\rm sys} = 0.50^{\prime\prime}$.}
The X-ray source in the vicinity of AT2024tvd is at
${\rm R.A.}=17^{\rm h}10^{\rm m}42.57^{\rm s}$, 
${\rm decl.}=+28^{\circ}50^{\prime}15.14^{\prime\prime}$.
Combining systematic and statistical errors, the $1\sigma$ uncertainty is \ad{$\sqrt{0.50^2+0.19^2}=0.53^{\prime\prime}$}.

Given the Chandra position, here we compute the positional posterior probabilities $P({\rm 24tvd|X})$ and $P({\rm nucleus|X})$ for the hypothesis that the X-ray source is associated with the optical transient AT2024tvd and the host galaxy nucleus. 
Let $P({\rm 24tvd})$ and $P({\rm nucleus})$ be the prior probabilities of the X-ray source being associated with 
AT2024tvd and the nucleus, respectively. To be conservative, we assume equal priors, $P({\rm 24tvd}) = P({\rm nucleus}) = 0.5$. 

We define a Cartesian coordinate system with the x-axis along the direction of right ascension, the y-axis along the direction of declination, and the origin $(0,0)$ being at the host galaxy nucleus (i.e., Figure~\ref{fig:offset}).
Therefore, the X-ray source is at coordinate $(x_0=0.485, y_0=0.829)$, and AT2024tvd is at coordinate (0.519, 0.752), as measured by HST.

The likelihood of the X-ray source being at a specific position is 
\begin{align}
    P({\rm X}|(x, y)) = \frac{1}{2\pi\sigma_x \sigma_y} {\rm exp}\left( -\frac{ r^2}{2  }\right),
\end{align}
where \ad{$\sigma_y = 0.53$, $\sigma_x = \sigma_y \cdot {\rm cos(decl.)}=0.46$}, and $r^2 = (x-x_0)^2/\sigma_x^2 +(y-y_0)^2/\sigma_y^2 $.

Applying Bayes' Theorem, we have 
\begin{align}
    & P({\rm 24tvd|X}) \notag \\
    =& \frac{P({\rm X| 24tvd}) P({\rm 24tvd})}{ P({\rm X| 24tvd}) P({\rm 24tvd}) + P({\rm X| nucleus}) P({\rm nucleus})  } \\
    =& \frac{P({\rm X| 24tvd}) }{ P({\rm X| 24tvd})  + P({\rm X| nucleus})  } =\ad{85}\% \notag ,
\end{align}
and $P({\rm nucleus|X}) = 1 - P({\rm 24tvd|X}) = \ad{15}$\%. Therefore, the X-ray source is most likely associated with the UV/optical transient AT2024tvd. We note that the deep eROSITA upper limit (Section~\ref{subsec:eROSITA}) and the rising XRT light curve (Section~\ref{subsec:xrt}) imply that the Chandra source is dominated by the transient, rather than being a blend of the transient and persistent emission from the host nucleus. 

\subsubsection{Chandra spectrum}

We extracted the source spectrum using a source region of $r_{\rm src}=1.5^{\prime\prime}$ centered on the X-ray position determined by \texttt{wavdetect} (see Section~\ref{subsubsec:cxo_astrometry}). A total of 24 counts were detected within the source region. The background spectrum was extracted using nearby source-free regions. We grouped the Chandra spectrum to at least one count per bin, and modeled the 0.3--7\,keV data using $W$-statistics. Using a model of \texttt{tbabs*zashift*ezdiskbb}, we obtained a poor fit with \texttt{cstat/dof=62.24/18}. 
\ad{Following Section~\ref{subsec:xrt}, we added an additional \texttt{simpl} component with $\Gamma$ fixed at 2.5}, which improves the fit to \texttt{cstat/dof=31.71/17}. 
The best-fit model gives \ad{a maximum disk temperature of $T_{\rm max}=95_{-12}^{+14}$\,eV, an upscattering fraction of $f_{\rm sc}=0.004_{-0.002}^{+0.004}$, and a 0.3--10\,keV flux of $2.6^{+1.4}_{-1.0}\times 10^{-13}\,{\rm erg\,s^{-1}\,cm^{-2}}$}.

\subsection{VLA} \label{subsec:vla}
Here we present a radio observation announced by \citet{Sfaradi2025_24tvd_radio_astronote}. A full analysis of a comprehensive radio follow-up campaign will be presented by Sfaradi et al. (2025, in preparation). 

We observed the field of AT2024tvd on 2025 January 3 ($\delta t = 105$\,d) using the Very Large Array (VLA; \citealt{Perley2011}) under Program 24A-386 (PI K. Alexander). The array was in its most extended A-configuration. 
We used the VLA calibration pipeline in the Common Astronomy Software Applications (\texttt{CASA}; \citealt{McMullin2007}) to flag and calibrate the data. J1735+3616 was used as an interleaved phase calibrator and 3C286 as the bandpass and absolute flux calibrator. The X-band image of the field around AT2024tvd was produced with the \texttt{CASA} task \texttt{tCLEAN}.
Our observation in X band (with a central frequency of 10\,GHz) results in an image rms of 10\,$\mu$Jy/beam.
The FWHM of the synthesized beam is $0.26^{\prime\prime}$ on the major axis and $0.21^{\prime\prime}$ on the minor axis, with the position angle being $-78^{\circ}$.
Flux uncertainties reported below have accounted for 10\% calibration uncertainties.

The lower right panel of Figure~\ref{fig:image} shows the radio image. 
A bright point source was detected near the phase center and we fitted it using CASA task IMFIT. 
Our best fit results in a flux density of $600 \pm 60\,\mu$Jy at 
${\rm R.A.}=17^{\rm h}10^{\rm m}42.571^{\rm s}$, 
${\rm decl.}=+28^{\circ}50^{\prime}15.064^{\prime\prime}$. 
This is $0.024^{\prime\prime}$ from the position of the HST TDE location (Section~\ref{subsubsec:hst_phot}).
For bright radio detections, the astrometric uncertainty is 10\% of the synthesized beam FWHM\footnote{The value of 10\% is recommended at \url{https://science.nrao.edu/facilities/vla/docs/manuals/oss/performance/positional-accuracy}.}, which is $0.024^{\prime\prime}$.
Therefore, this source is consistent with being the radio counterpart of AT2024tvd. 

We also detect a $50 \pm 11\,\mu$Jy source at 
${\rm R.A.}=17^{\rm h}10^{\rm m}42.542^{\rm s}$, 
${\rm decl.}=+28^{\circ}50^{\prime}14.293^{\prime\prime}$.
However, we note that this is only a $5\sigma$ detection and the source cannot be well described by a point source (see the contours in Figure~\ref{fig:image}).
Therefore, although it is $0.12^{\prime\prime}$ from the HST host galaxy center (Section~\ref{subsubsec:hst_phot}), greater than the nominal VLA astrometric uncertainty, we still consider it to be likely associated with the host galaxy nucleus. 

\section{Discussion}
\label{sec:discuss}

\subsection{An Off-nuclear TDE}

AT2024tvd exhibits all hallmark properties of previously known nuclear TDEs.
Its UV and optical emission remains hot ($T_{\rm bb}\sim 2\times 10^4$\,K) throughout the evolution (Section~\ref{subsec:bbfit}).
\ad{Its UV spectrum bears a remarkable resemblance to that of the TDE ASASSN-14li (Section~\ref{subsubsec:hst_spec}).}
Its optical spectra exhibit broad hydrogen lines (Section~\ref{subsec:opt-spec}).
Observations with Chandra provide evidence for the physical association of the X-ray source with the UV/optical TDE (Section~\ref{subsec:chandra}).
The X-ray emission is luminous ($L_{\rm X} \sim 10^{43}\,{\rm erg\,s^{-1}}$) and soft, with $T_{\rm in}\sim 0.1$--0.2\,keV or $3<\Gamma<6$ (Section~\ref{subsec:xrt}). This is typical for TDEs (see, e.g., \citealt{Guolo2024} and Tab.~1 of \citealt{Saxton2020}). 
Significant hour-timescale X-ray variability has been observed, similar to previously known TDEs such as 2MASX\,0740-85 \citep{Saxton2017} and AT2022lri \citep{Yao2024}.
At a phase of $\delta t = 105$\,d, the radio luminosity of AT2024tvd is $L_{10\,\rm GHz}\sim 3\times 10^{38}\,{\rm erg\,s^{-1}}$ (Section~\ref{subsec:vla}), similar to the radio luminosity of some known radio-bright non-jetted TDEs, such as ASASSN-15oi \citep{Horesh2021_15oi}, AT2019dsg \citep{Stein2021, Cendes2021_19dsg}, and AT2020opy \citep{Goodwin2022}.

We note that a supernova origin is confidently ruled out. In the majority of supernovae (SNe), the UV and optical emission significantly cool over time as the photosphere expands. The only types of SNe that may remain hot around peak light are hydrogen-poor superluminous supernovae (SLSNe I), and those powered by interaction between the outgoing ejecta and the ambient circumstellar medium (CSM), including SNe IIn/Ibn/Icn, hydrogen-rich superluminous supernovae (SLSNe II), and SNe Ia-CSM. However, even the superluminous/interacting SNe still cool down to temperatures much less than $10^4$\,K by a couple of months post-peak \citep{Gomez2024, Ransome2024}. Around 5000\,K is a typical temperature at that phase, well below that observed in AT2024tvd.
In AT2024tvd, the X-ray luminosity of $\sim 10^{43}\,{\rm erg\,s^{-1}}$ is more luminous than any known SN (see, e.g., compilations in Fig.~1 of \citealt{Dwarkadas2014} and Fig.~5 of \citealt{AXIStdamm2024}). The X-ray spectrum is also significantly softer than normal supernova spectra and not consistent with interaction.

Five instruments (P48/ZTF, Swift/UVOT, Chandra, VLA, and HST) independently demonstrate and support that the TDE location is offset from the galaxy nucleus. Among them, HST and VLA spatially resolved the emission from the galaxy nucleus \emph{and} the transient (Figure~\ref{fig:image}). The separation is measured to be $0.914\pm0.010$\arcsec\ by HST, which corresponds to a projected physical distance of $0.808\pm0.009$\,kpc.

A radio source with $L_{\rm 10\,GHz}=2.4\times 10^{37}\,{\rm erg\,s^{-1}}$ is detected at the galaxy centroid (Section~\ref{subsec:vla}). If this is powered by star formation, using the star formation rate (SFR) versus $L_{\rm 1.4\,GHz}$ relation from \citet{Davies2017} and assuming a typical spectral shape of $f_{\nu}\propto \nu^{-0.8}$ \citep{Magnelli2015}, 
the SFR needs to be $0.7\,M_\odot\,{\rm yr^{-1}}$.
However, stellar population synthesis analysis constrains the SFR of the host galaxy to be $<0.1\,M_\odot\,{\rm yr^{-1}}$ \citep{Maraston2009}. Therefore, this radio source must be powered by a low-luminosity AGN, suggesting the existence of at least one MBH in the galaxy nucleus. The inferred MBH mass in the galaxy center is $M_{\rm BH} \sim 10^{8.4}\,M_\odot$ (Section~\ref{subsec:sdss_spec}). 

Using the X-ray to radio luminosity correlation for low-luminosity radio galaxies \citep{Panessa2007}, the expected X-ray luminosity is $\sim10^{38.7}\,{\rm erg\,s^{-1}}$, consistent with the eROSITA upper limit (Section~\ref{subsec:eROSITA}).
Assuming a typical radio spectral shape of $f_{\nu}\propto \nu^{-0.63}$ \citep{Sabater2019}, we calculate a radio luminosity of $L_{\rm 150 \, MHz} \sim  3\times 10^{21}\,{\rm W\, Hz^{-1}}$.
Such a luminosity is expected in massive galaxies with $M_{\rm gal}\sim 10^{11}\,M_\odot$ 
\citep{Sabater2019}. 

Using the stellar--halo mass relation \citep{Moster2013}, we estimate that the halo mass is $M_{\rm h}\sim 10^{13}\,M_\odot$ for the host galaxy mass of $\sim 10^{11}\,M_\odot$. The virial radius is $R_{200}=424$\,kpc.

We conclude that AT2024tvd is the first off-nuclear TDE selected from optical sky surveys. 

\begin{figure*}[htbp!]
    \centering
    \includegraphics[width=\textwidth]{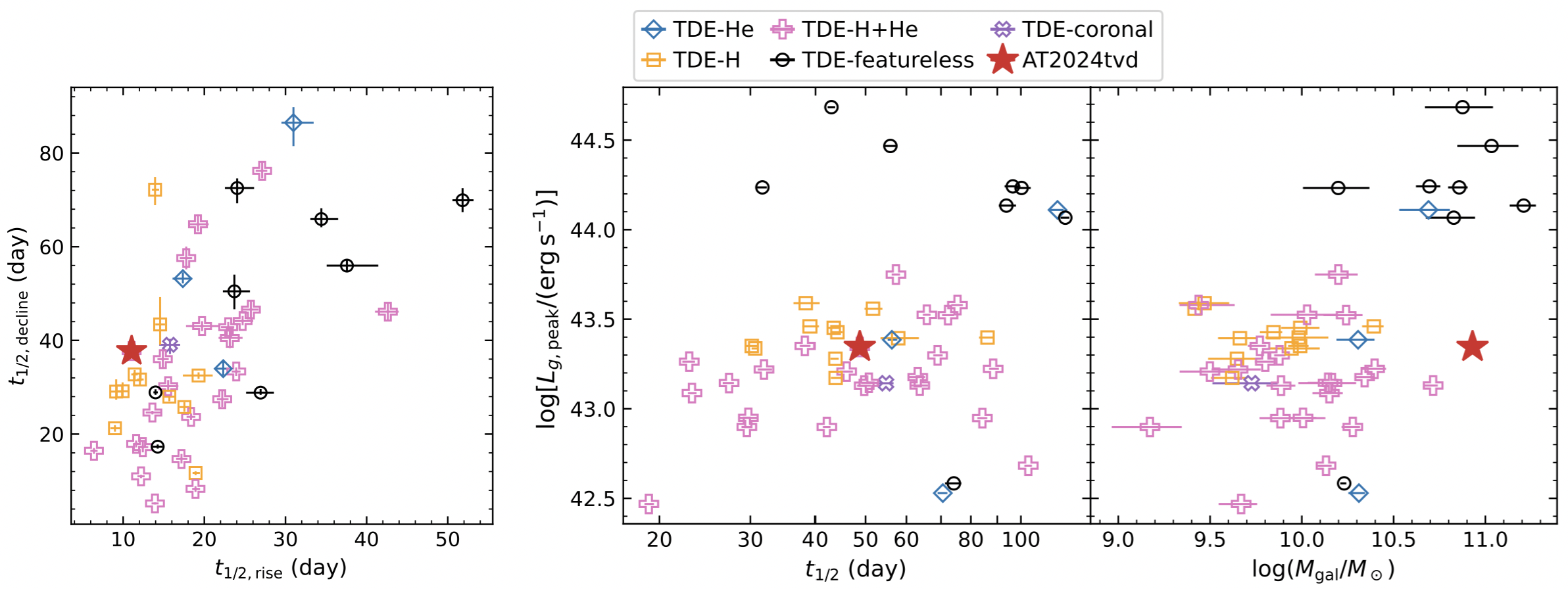}
    \caption{Light curve and host galaxy properties of AT2024tvd compared with published ZTF TDE sample, color-coded by the optical spectral subtype \citep{vanVelzen2021, Hammerstein2023, Yao2023}.\label{fig:24tvd_compare}
    }
\end{figure*}

\subsection{Comparison with Known TDEs}
In Figure~\ref{fig:24tvd_compare}, we compare the light curve peak rest-frame $g$-band luminosity ($L_{g, {\rm peak}}$), rest-frame duration above half-maximum ($t_{1/2}$), rise time from half-maximum to maximum ($t_{1/2, {\rm rise}}$), decline time from maximum to half-maximum ($t_{1/2, \rm decline}$), and host galaxy total stellar mass ($M_{\rm gal}$) of AT2024tvd with those of 45 previously known ZTF TDEs. The comparison sample is compiled from ZTF TDEs with identified optical spectral subtypes presented by \citet{vanVelzen2021}, \citet{Hammerstein2023}, and \citet{Yao2023}.

The left and middle panels of Figure~\ref{fig:24tvd_compare} display the light curve parameters, derived using the method outlined in \citet{Yao2023}. For AT2024tvd, we find 
$L_{g, {\rm peak}}=10^{43.35}\,{\rm erg\,s^{-1}}$, 
$t_{1/2} = 48.7_{-1.1}^{+0.9}$\,d, 
$t_{1/2, \rm rise} = 11.0_{-0.5}^{+0.4}$\,d, and 
$t_{1/2, \rm decline} = 37.7_{-1.0}^{+0.8}$\,d.
Among the comparison sample, 33 TDEs have broad hydrogen lines (i.e., the TDE-H+He and TDE-H subtypes), with median values of 
$L_{g, {\rm peak}}=10^{43.26}\,{\rm erg\,s^{-1}}$, 
$t_{1/2} = 44.1$\,d, 
$t_{1/2, \rm rise} = 17.2$\,d, and 
$t_{1/2, \rm decline} = 30.2$\,d.
The UV/optical light curve properties of AT2024tvd align well with the typical characteristics of the TDE-H+He and TDE-H subtypes, although its rise time is on the fast side.

The right panel of Figure~\ref{fig:24tvd_compare} displays the events on the $L_{g, {\rm peak}}$ versus $M_{\rm gal}$ diagram. Among the 33 events classified as TDE-H+He or TDE-H subtypes, the median value of their host galaxy stellar mass is $10^{9.9}\,M_\odot$, with all events occurring in galaxies less massive than AT2024tvd’s host.
AT2024tvd also appears to be underluminous compared with other TDEs hosted by galaxies with $M_{\rm gal}\sim 10^{11}\,M_\odot$. 
This suggests the MBH in AT2024tvd is probably of much lower mass than the $M_{\rm BH}\sim 10^{8}\,M_\odot$ black holes typically found in the centers of $\sim 10^{11}\,M_\odot$ galaxies. 

Several approaches to infer $M_{\rm BH}$ based on TDE observables have been proposed, including (1) fitting the UV and optical light curves \citep{Mockler2019}, (2) using the peak $g$-band luminosity or the luminosity of a late-time UV plateau \citep{Mummery2024_fundamental_scaling_relation}, and (3) modeling of the X-ray spectra \citep{Wen2021, Guolo2025}. 
While none of these approaches have been extensively tested, we apply methods (1) and (2) to estimate the black hole mass of AT2024tvd. 
First, we use the TDE module of Modular Open-Source Fitter for Transients (\texttt{MOSFiT}; \citealt{Guillochon2018, Mockler2019}). In \texttt{MOSFiT}, the mass fall-back rate ($\dot M_{\rm fb}$) is constructed based on hydrodynamical simulations; the mass accretion rate is viscously delayed relative to $\dot M_{\rm fb}$; the UV/optical is assumed to be blackbody emission generated in a reprocessing region; and the photospheric radius has a power-law dependence on the luminosity.
Our fitting result gives a black hole mass of ${\rm log}(M_{\rm BH}/M_\odot) = 5.89_{-0.06}^{+0.15}$.
Next, using Eq.~(74) of \citet{Mummery2024_fundamental_scaling_relation}, we obtain ${\rm log}(M_{\rm BH}/M_\odot) =6.9\pm0.5$.

\begin{figure}[htbp!]
    \centering
    \includegraphics[width=\columnwidth]{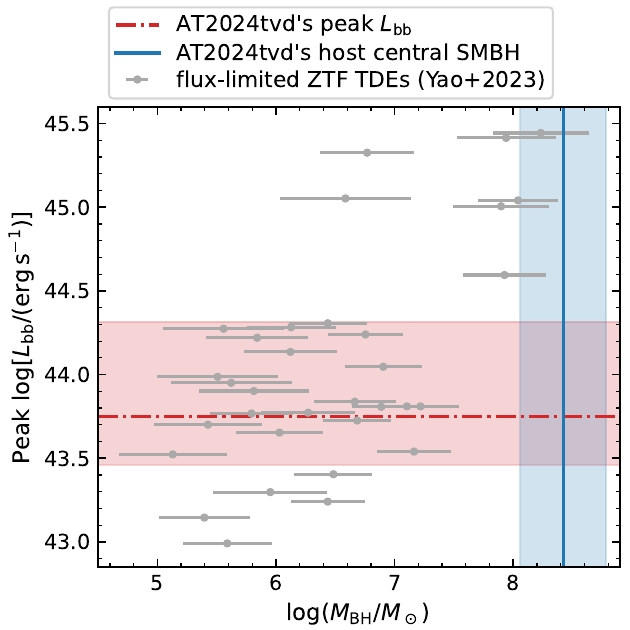}
    \caption{The grey data points show the peak luminosity of the UV/optical emission component versus host galaxy nucleus black hole mass for a flux-limited sample of 33 ZTF TDEs \citep{Yao2023}. Comparing the peak $L_{\rm bb}$ observed in AT2024tvd (the horizontal dash-dotted red line) with the known sample, the black hole mass in AT2024tvd is likely between $10^{5.1}\,M_\odot$ and $10^{7.2}\,M_\odot$ --- more than an order of magnitude lower than the $M_{\rm BH}$ of the black hole in the center of the host galaxy (the vertical blue line). \label{fig:Lbb_Mbh}}
\end{figure}

The peak bolometric luminosity of the UV/optical emission of AT2024tvd is ${\rm log} [L_{\rm bb}/({\rm erg\,s^{-1}})] = 43.75_{-0.27}^{+0.57}$ (Section~\ref{subsec:bbfit}). 
Figure~\ref{fig:Lbb_Mbh} shows that among the flux-limited sample of ZTF TDEs in \citet{Yao2023} with such a peak luminosity, their host MBH mass is between $10^{5.1}\,M_\odot$ and $10^{7.2}\,M_\odot$, with the median being $10^{6.1}\,M_\odot$. We therefore deduce that the MBH mass of AT2024tvd is likely ${\rm log}(M_{\rm BH}/M_\odot)=6\pm1$. This is consistent with our results derived with \texttt{MOSFiT} and the \citet{Mummery2024_fundamental_scaling_relation} scaling relation. 

\subsection{Formation Scenarios} \label{subsec:scenarios}

\begin{deluxetable*}{ccccccc}[htbp!]
    \tablecaption{Summary of off-nuclear TDEs.\label{tab:off-nuclear_tde}}
	\tablehead{
	\colhead{Name}
		& \colhead{$z$} 
		& \colhead{Offset}
        & \colhead{Parent $M_{\rm gal}$} 
        & \colhead{Satellite dwarf $M_{\ast}$} 
        & \colhead{Central $M_{\rm BH}$} 
        & \colhead{TDE $M_{\rm BH}$} 
        \\
        \colhead{}
        & \colhead{}
        & \colhead{(\arcsec; kpc)}
        & \colhead{($M_\odot$)}
        & \colhead{($M_\odot$)}
        & \colhead{($M_\odot$)}
        & \colhead{($M_\odot$)}
	}
	\startdata
       3XMM\,J2150  & 0.05526 & 11.6; 12.5  & $10^{10.93\pm0.07}$ & $10^{7.3\pm0.4}$ & \ad{$10^{8.49\pm0.67}$} & $\sim 10^{4.9}$\\
       EP240222a  &  0.03275 & 53.1; 34.7   & $10^{10.89\pm0.07}$ & $10^{7.0\pm0.3}$ & \ad{$10^{8.44\pm0.67}$} & $\sim 10^{4.9}$ \\
       AT2024tvd  &  0.04494 & 0.92; 0.81   & $10^{10.93\pm0.02}$ & \ad{$<10^{7.6}$}            & \ad{$10^{8.37\pm0.51}$} & $\sim 10^{5.9}$\\
    \enddata
    \tablecomments{The TDE $M_{\rm BH}$ estimates come from fitting to the X-ray spectra (for 3XMM\,J2150 and EP240222a) or the UV/optical light curve (AT2024tvd), while we note that both methods are subject to some uncertainties. }
\end{deluxetable*}

There are three possible scenarios for offset TDEs:
\begin{enumerate}
    \item The TDE originates from one black hole in a dual or inspiraling pre-merger MBH system.
    \item The disruption is produced by an ejected MBH from a triple system (gravitational slingshot).
    \item The TDE occurs due to a recoiling MBH kicked by the GW emission following MBH coalescence (gravitational rocket). 
\end{enumerate}
Scenario 3 is very unlikely because (1) the existence of a radio source at the galaxy nucleus indicates the existence of at least one MBH, and (2) the inferred mass of AT2024tvd's MBH is too small to agree with the total stellar mass of its host galaxy. Below we discuss the other two viable scenarios.

\subsubsection{A MBH pair from a galaxy minor merger}
Scenario 1 provides a natural explanation, where the MBH of AT2024tvd originates from the center of a galaxy destroyed in a minor merger. 
Since the DF timescale can exceed the Hubble time for lower-mass secondary MBHs \citep{Dosopoulou2017, Kelley2017}, we expect high fractions of wandering $<10^8\,M_\odot$ MBHs from the cumulative merger history.
Simulations have shown that there can be significant time delays between galaxy merger and black hole merger, even on Galactic scales \citep{Tremmel2018_apjl}. 

In Table~\ref{tab:off-nuclear_tde}, we compare the properties of the three off-nuclear TDEs known to date. 
Using the method outlined in Section~\ref{subsec:host_sed}, we estimate the total stellar mass of the lenticular galaxy 12.5\,kpc from 3XMM\,J2150 to be ${\rm log}(M_{\rm gal}/M_\odot) = 10.93_{-0.08}^{+0.06}$, and its central black hole mass to be \ad{${\rm log}(M_{\rm BH}/M_\odot) = 8.49\pm0.67$}. 
Similarly, for the parent galaxy of EP240222a, we have ${\rm log}(M_{\rm gal}/M_\odot) = 10.89_{-0.08}^{+0.07}$ \citep{Chang2015} and \ad{${\rm log}(M_{\rm BH}/M_\odot) = 8.44\pm0.67$}. 

We see that the parent galaxies of all three off-nuclear TDEs are very massive, with $M_{\rm gal}\sim 10^{10.9}\,M_\odot$.
This is consistent with cosmological simulations by \citet{Ricarte2021_wanderers_origin}, which predicts that the number of wanderers\footnote{Whether or not the BHs associated with 3XMM\,J2150 and EP240222a can be called `wanderers' is a grey area. Although both are still bound to their dwarf hosts, the amount of bound material (stellar and dark matter) will gradually decrease over time.} grows linearly with halo mass. 
At a halo mass of $M_{\rm h}\sim 10^{13}\,M_\odot$ and at $z=0.05$, the average number of wanderers is $\sim 10^2$, and about 10\% of these wanderers have $M_{\rm BH}>10^6\,M_\odot$. 
Off-nuclear TDEs have not been found in even higher mass galaxies, probably because the present-day galaxy stellar mass function exhibits an exponential cut-off above $10^{10.8}\,M_\odot$ \citep{Wright2017}.

The optical sources/satellites associated with 3XMM\,J2150 and EP240222a, with stellar masses of $\sim 10^{7}\,M_\odot$ \citep{Lin2018_3XMMJ2150, Jin2025}, are spatially resolved, whereas we do not detect any residual features at the position of AT2024tvd in archival optical images (Section~\ref{subsec:archival_image_analysis}). This is also consistent with results from \citet{Ricarte2021_wanderers_origin}, showing that wandering MBHs in stellar overdensities tend to exist only at larger halo-centric radii where the tidal forces are weaker, and the secondary galaxy is able to retain more material. 

\subsubsection{An ejected MBH from a triple system}

In scenario 2, the ejected MBH must have a velocity that's greater than the stellar velocity dispersion of $\sim200\,{\rm km\,s^{-1}}$ and comparable to the Galactic escape velocity of $\sim10^3\,{\rm km\,s^{-1}}$, otherwise the ejected BH remains in the proximity of the galaxy center \citep{Hoffman2006}.
Simulations show that most of the single MBHs are ejected to the outskirts of or become unbound to the host galaxies \citep{Volonteri2005_imbh_dynamics}. 
In those ejected cases, the total time spent by the single MBH at a separation of $\sim 0.8$\,kpc from the nucleus is $<4$\,Myr. 
If not, the MBH will oscillate in the galactic potential as their orbits decay by DF \citep{Madau2004, Gualandris2008}. The time that it spends (from Myr to Gyr) and the maximum distance reached before returning to the center largely depends on the ejection velocity and central density gradients \citep{Stone2012, Blecha2016}. For typical bound systems modeled by \citet{Hoffman2007}, the recoiling MBH spends $\sim 10$\,Myr at $\sim 0.8$\,kpc. 


While no theoretical studies have specifically examined TDE rates from recoiling MBHs following a slingshot, relevant insights can be drawn from  \citet{Komossa2008} and \citet{Stone2012}. These studies carried out calculations for TDE rates from recoiling MBHs kicked by GW, finding that the recoiling MBH will carry a star cluster with it. At a distance of $\sim 0.8$\,kpc from the nucleus, the TDE rate will be dominated by the bound stars, producing TDEs at a rate of $\sim1$\% the nuclear TDE rate. 
Given that ZTF has found $\sim10^2$ TDEs, it is possible that AT2024tvd originates from an off-nuclear MBH formed through this mechanism. Future detailed modeling of the host galaxy is needed to determine whether the timescale on which stars in the bound cluster are depleted is long enough to sustain such an event rate.

\subsubsection{Comparison with dual and offset AGN at $\lesssim1$\,kpc}
The search for MBH pairs via dual and offset AGN has been an effort for over four decades. 
Nonetheless, at sub-kiloparsec separations, fewer than ten dual AGN have been identified (see recent summaries in \citealt{Chen2022_dual_agn} and \citealt{Puerto-Sanchez2025}), and \ad{only one} offset single-AGN has been confirmed \ad{\citep{Schweizer2018, Voggel2022}}. 

The most well-studied dual AGN are those in the nearby merging galaxy NGC\,6240 \citep{Komossa2003, Medling2011, Muller-Sanchez2018} and UGC\,4211 \citep{Koss2023}.
Both systems exhibit heavily disturbed morphologies, prominent large-scale tidal features and dust lanes, all consistent with ongoing gas-rich major mergers.
\ad{Similarly, the offset AGN in NGC 7727 is associated with a long blue tidal stream \citep{Schweizer2018}.}
In contrast, the host galaxy of AT2024tvd appears relatively undisturbed, lacking visible tidal tails or arcs. 
This is similar to the three galaxies at $z\sim0.1$ hosting dual AGN reported by \citet{Muller-Sanchez2015}. 
Such systems can be explained by a minor merger where the primary galaxy is relatively undisturbed, a major merger that occurred too long ago for the system's dynamics to have relaxed, or a combination of the two. 
For AT2024tvd, the large mass ratio between the black hole in the galaxy's center ($M_{\rm BH}\gtrsim10^8\,M_\odot$) and the black hole powering the transient ($10^5\,M_\odot < M_{\rm BH}< 10^8\,M_\odot$) suggests that at least a minor merger plays a role.

\section{Summary and Future Perspective}
\label{sec:summary}

We presented AT2024tvd, a TDE offset by $0.914\pm 0.010$\arcsec\ from its host galaxy’s nucleus, corresponding to a projected distance of $0.808\pm 0.009$\,kpc. The inferred mass of the black hole powering AT2024tvd is in the range of $10^5$--$10^7\,M_\odot$, at least an order of magnitude smaller than the supermassive black hole at the center of its host galaxy. 

AT2024tvd represents the first off-nuclear TDE identified through optical sky surveys. It likely originates from either (i) a minor galaxy merger, where the TDE occurs in the least massive galaxy during the dynamical friction phase, or (ii) a recoiling MBH ejected via a slingshot in triple MBH interactions. 
In both scenarios, a surrounding star cluster is needed to supply TDEs. 
Archival optical images constrain the mass of any such star cluster to \ad{$M_\ast < 10^{7.6}\,M_\odot$ (Section~\ref{subsec:archival_image_analysis})}. 
Deeper constraints on the stellar counterpart bound to AT2024tvd’s MBH may be achieved with future HST or JWST observations once the TDE emission fades.

Unlike the two previously known off-nuclear TDEs, which are linked to disrupted satellite dwarf galaxies in the outskirts of their parent galaxies, AT2024tvd lies well within the galactic bulge of its host galaxy.
The total stellar mass of the parent galaxies of all three off-nuclear TDEs is $10^{10.9}\,M_\odot$. Under the picture of scenario (i), this is consistent with cosmological simulations that massive halos host more wandering black holes. 
Due to event horizon suppression, the nuclear TDE rate exhibits a sharp drop off above $M_{\rm gal}\sim 5\times 10^{10}\,M_\odot$ or $M_{\rm BH}\sim 10^8\,M_\odot$ \citep{Yao2023}. Therefore, future searches for offset MBHs could efficiently start with TDEs in massive galaxies.

TDEs are an incredibly valuable probe of dual MBH systems. Compared with AGN, TDEs are subject to very different selection criteria. Particularly, they can probe the otherwise quiescent systems, MBH masses that are relatively low, and conceivably dual MBH's where the two MBHs cannot be spatially resolved. In the final case, it would be a challenge to demonstrate a dual system, but possible with kinematic signatures and arguments based on MBH masses. 

Development of photometric TDE selection filters that are agnostic of proximity to a cataloged galaxy nucleus is needed to efficiently identify off-nuclear TDE candidates. 
We anticipate a growing number of off-nuclear TDE discoveries with future sky surveys, such as the Legacy Survey of Space and Time (LSST) at the Vera C. Rubin Observatory. With an unprecedented sensitivity ($r\sim 24.5$) and an astrometric precision of 10\,mas \citep{Ivezic2019}, LSST can uncover off-nuclear TDEs out to cosmological distances. These discoveries will significantly advance our understanding of the formation, dynamics, and demographics of off-nuclear MBHs and their transient activity.

\begin{acknowledgments}

\ad{We thank the anonymous referee for constructive comments and suggestions.}
YY would like to thank Angelo Ricarte, \ad{Anil Seth}, and Nick Stone for helpful conversations about the origin \ad{and search} of offset MBHs.
CW would like to thank Peter Melchior for helpful discussions about running \texttt{Scarlet} on HST imaging. 
We thank Kirsty Taggart for assistance with the Kast data reduction. 

CW acknowledges support from the LSST Discovery Alliance under grant AWD1008640. 
RM acknowledges support by the National Science
Foundation under award No. AST-2224255.
CL, AAM and NR are supported by DoE award \#DE-SC0025599. 
KDA acknowledges support provided by the NSF through award AST-2307668.  
MN is supported by the European Research Council (ERC) under the European Union’s Horizon 2020 research and innovation programme (grant agreement No.~948381) and by UK Space Agency Grant No.~ST/Y000692/1.

Based on observations obtained with the Samuel Oschin Telescope 48-inch and the 60-inch Telescope at the Palomar Observatory as part of the Zwicky Transient Facility project. ZTF is supported by the National Science Foundation under Grant No. AST-2034437 and a collaboration including Caltech, IPAC, the Oskar Klein Center at Stockholm University, the University of Maryland, University of California, Berkeley, the University of Wisconsin at Milwaukee, University of Warwick, Ruhr University Bochum, Cornell University, Northwestern University and Drexel University. Operations are conducted by COO, IPAC, and UW. 
SED Machine is based upon work supported by the National Science Foundation under Grant No. 1106171. 
The Gordon and Betty Moore Foundation, through both the Data-Driven Investigator Program and a dedicated grant, provided critical funding for SkyPortal.
The ZTF forced-photometry service was funded under the Heising-Simons Foundation grant No. 12540303 (PI: Graham).

This work has made use of data from the Asteroid Terrestrial-impact Last Alert System (ATLAS) project. The ATLAS project is primarily funded to search for near earth asteroids through NASA grants NN12AR55G, 80NSSC18K0284, and 80NSSC18K1575; byproducts of the NEO search include images and catalogs from the survey area. This work was partially funded by Kepler/K2 grant J1944/80NSSC19K0112 and HST GO-15889, and STFC grants ST/T000198/1 and ST/S006109/1. The ATLAS science products have been made possible through the contributions of the University of Hawaii Institute for Astronomy, the Queen’s University Belfast, the Space Telescope Science Institute, the South African Astronomical Observatory, and The Millennium Institute of Astrophysics (MAS), Chile.

The scientific results reported in this article are based on observations made by the Chandra X-ray Observatory.
This work uses data obtained with eROSITA telescope onboard SRG observatory. The SRG observatory was built by Roskosmos with the participation of the Deutsches Zentrum für Luft- und Raumfahrt (DLR). The SRG/eROSITA X-ray telescope was built by a consortium of German Institutes led by MPE, and supported by DLR. The SRG spacecraft was designed, built, launched and is operated by the Lavochkin Association and its subcontractors. The science data were downlinked via the Deep Space Network Antennae in Bear Lakes, Ussurijsk, and Baykonur, funded by Roskosmos. The eROSITA data used in this work were processed using the eSASS software system developed by the German eROSITA consortium and proprietary data reduction and analysis software developed by the Russian eROSITA Consortium.

MMT Observatory and Zwicky Transient Facility access was supported by Northwestern University and the Center for Interdisciplinary Exploration and Research in Astrophysics (CIERA).

The National Radio Astronomy Observatory is a facility of the National Science Foundation operated under cooperative agreement by Associated Universities, Inc.

\end{acknowledgments}

\facilities{
PO:1.2m, 
PO:1.5m, 
Shane,
MMT,
Swift,
EVLA,
CXO,
ATLAS, 
VLA
}

\software{
\texttt{astropy} \citep{Astropy2013}, \texttt{CASA} \citep{McMullin2007},
\texttt{emcee} \citep{Foreman-Mackey2013}, 
\texttt{heasoft} \citep{HEASARC2014}, 
\texttt{matplotlib} \citep{Hunter2007}, 
\texttt{Scarlet} \citep{Melchior2018}, 
\texttt{Scarlet2} \citep{Sampson2024, Ward2024}, 
\texttt{scipy} \citep{Virtanen2020}, 
\texttt{xspec} \citep{Arnaud1996}.
}

\appendix

\section{Multi-epoch ZTF image modeling to confirm the transient--host offset} \label{sec:ztf_astrometry}

With the \texttt{Scarlet2} framework, we were able to include pre-flare ZTF imaging to constrain the host galaxy model, and jointly model the position and flux of the transient in the images containing the TDE. We used a non-parametric model for the host galaxy with the \texttt{ZTF\_ScoreNet32} prior, which was trained on low redshift ZTF host galaxies \citep{Sampson2024}, to obtain an optimal fit for the position of the transient. We then applied a parametric model to the host galaxy in pre-flare imaging to extract its center. In this way, we were able to ensure that the galaxy center as measured in the ZTF images was consistent with the center measured from LS and PS1 imaging, regardless of astrometric mismatches, and measure the transient-host offset using consistent imaging data instead of comparing positions across surveys.

\begin{figure*}[htbp!]
\gridline{\fig{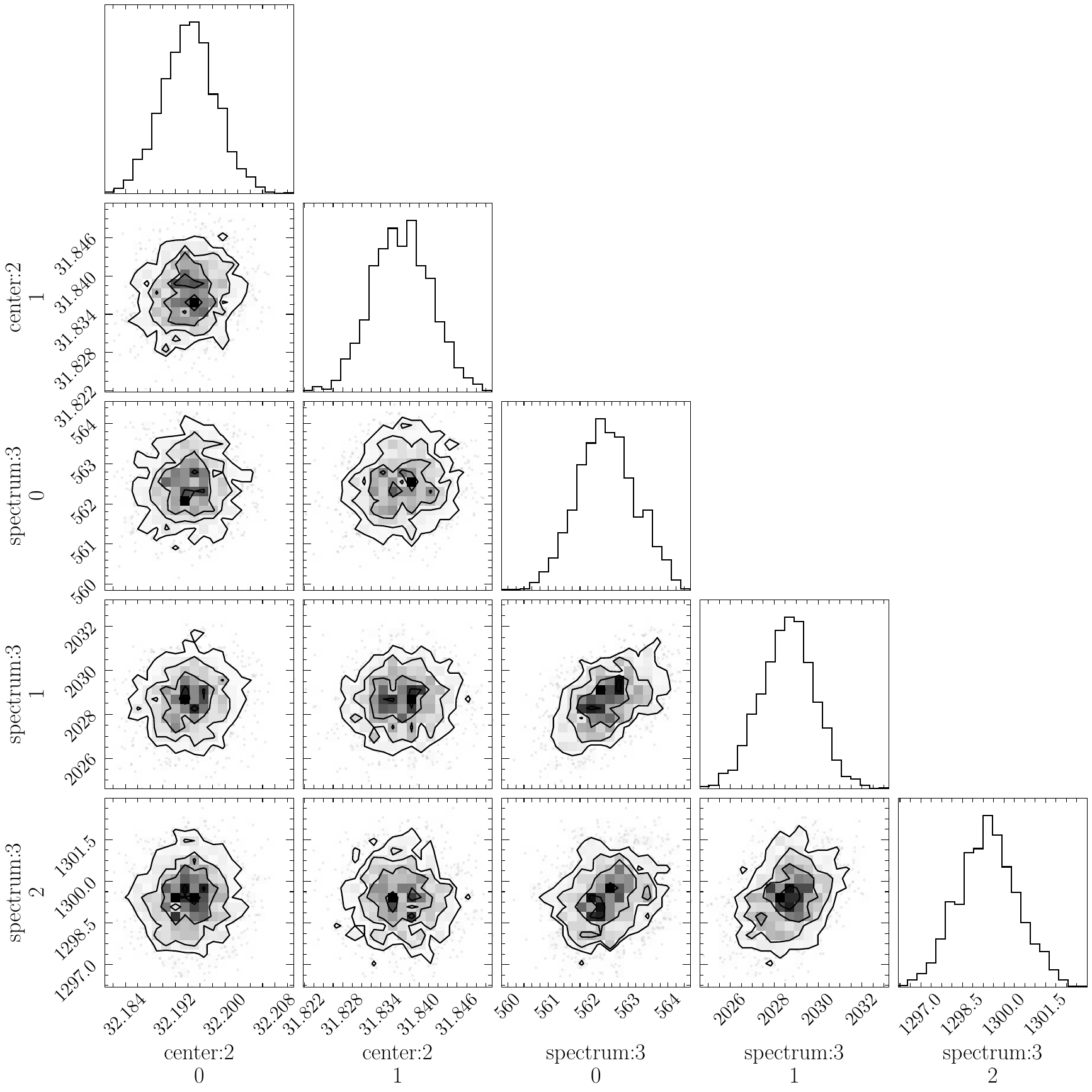}{0.6\textwidth}{}} 
\caption{\textit{Top}: Posteriors from sampling over the variable point source and host galaxy parameters in the multi-epoch ZTF imaging. We only show posteriors for the TDE position and the host galaxy SED, as there are 41 free TDE flux parameters for each epoch, and 900 free `parameters' for each pixel in galaxy morphology model of box size 30 by 30.}
\label{fig:PosteriorTDE}
\end{figure*}

\begin{figure*}[htbp!]
\gridline{\fig{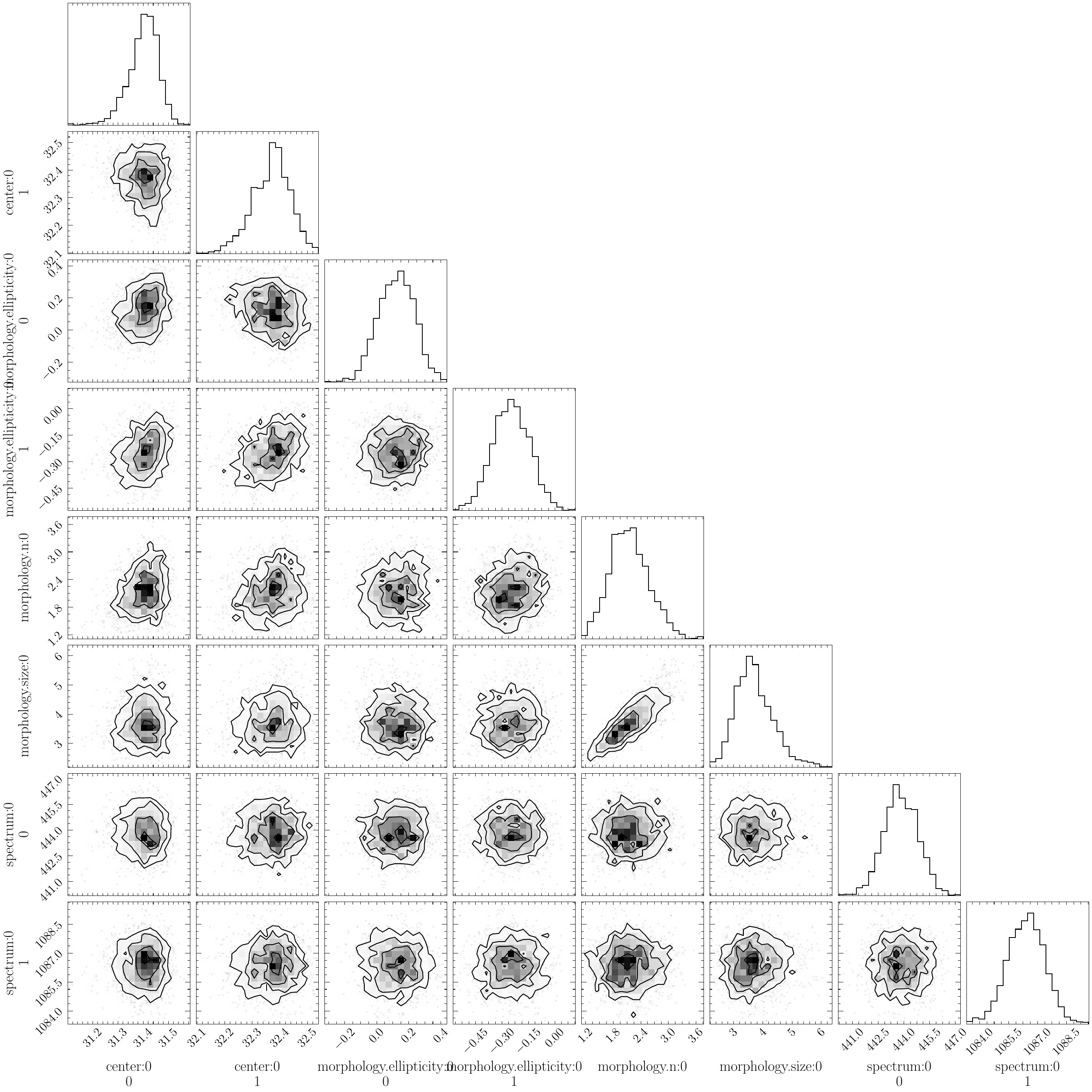}{0.8\textwidth}{}} 
\caption{Posteriors from sampling over host galaxy parameters from the stack of pre-flare ZTF images.}
\label{fig:Posteriorgalaxymodel}
\end{figure*}

We first used the \texttt{ZTFquery} cutout service \citep{Rigault2018} to download $120^{\prime\prime}$ by $120^{\prime\prime}$ cutouts of the $g$ and $r$-band  ZTF single-epoch imaging. We required that the images have seeing $<2^{\prime\prime}$ and limiting magnitude $>20$, and selected 21 pre-flare images prior to MJD 60537 and 21 images when the TDE was present. We ran the wavelet detection routine implemented in \texttt{Scarlet} on the summed images using the first three wavelet levels to find positions of all sources in the coadd that were detected at $>5\sigma$. We initialized extended sources at each position. At the position of the TDE host galaxy identified by the wavelet detection routine, we also initialized a variable point source at that same position. Repeating the procedure described in \citet{Ward2024} for the ZTF TDE host galaxies, we constrained each source to have positive flux and morphology models, and required that the TDE flux be zero in images from ${\rm MJD}<60537$ to remove degeneracies with the host model. We fit the scene until a relative error of $10^{-6}$ was reached, or a maximum of 3000 steps. After obtaining the scene model, we used the \texttt{numpyro} NUTS MCMC sampling routine \citep{Phan2019, Bingham2019} implemented within \texttt{Scarlet2} to sample over the point source position. The fitting procedure identified the position of the transient and its $3\sigma$ error to be ${\rm R.A.} (^{\circ})=257.6773859\pm0.0000034$ and ${\rm decl.} (^{\circ})= +28.8375471\pm0.0000037$. The posteriors for the galaxy parameters are shown in Figure \ref{fig:PosteriorTDE}.

We then produced a stack of the 21 pre-flare ZTF images to produce a high S/N multi-band image of the host galaxy. We fit a S\'{e}rsic galaxy model to the stack to fit the center of the galaxy by initializing a S\'{e}rsic profile and allowing \texttt{Scarlet2} to fit the half light radius, ellipticity, S\'{e}rsic index, spectrum, and central position. The MCMC sampling routine was again applied to determine the galaxy model posteriors. The fitting procedure identified the position of the host galaxy center and its $3\sigma$ error to be ${\rm R.A.} (^{\circ})=257.67721\pm0.00012$ and ${\rm decl.} (^{\circ})= 28.83733\pm0.00011$ under the assumption of a S\'{e}rsic profile, such that the offset is $0.95\pm0.42^{\prime\prime}$ (3$\sigma$ uncertainty). The posteriors for the galaxy parameters are shown in Figure \ref{fig:Posteriorgalaxymodel}.

The code used to perform the \texttt{Scarlet2} analysis described in this Section is available at \url{https://github.com/yaoyuhan/24tvd_discovery_paper/blob/main/offset/ZTF22aaigqsr_scrlet2_analysis_example.ipynb}.

\section{Additional Figures} \label{sec:figures}

\begin{figure*}[htbp!]
\gridline{\fig{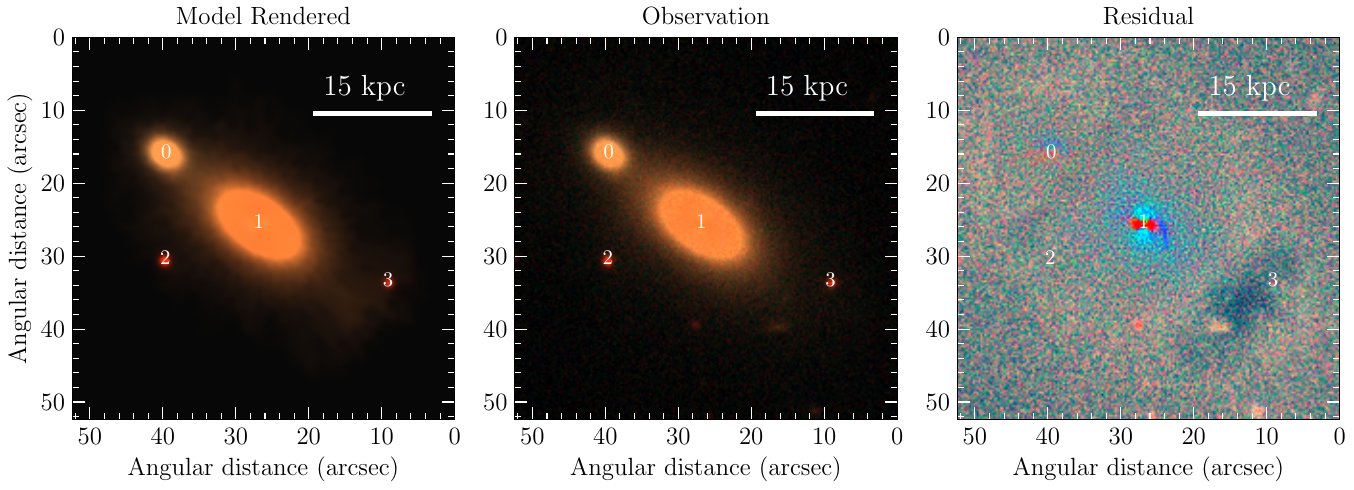}{0.7\textwidth}{Legacy Survey DR10 imaging ($grz$)}} 
 \gridline{\fig{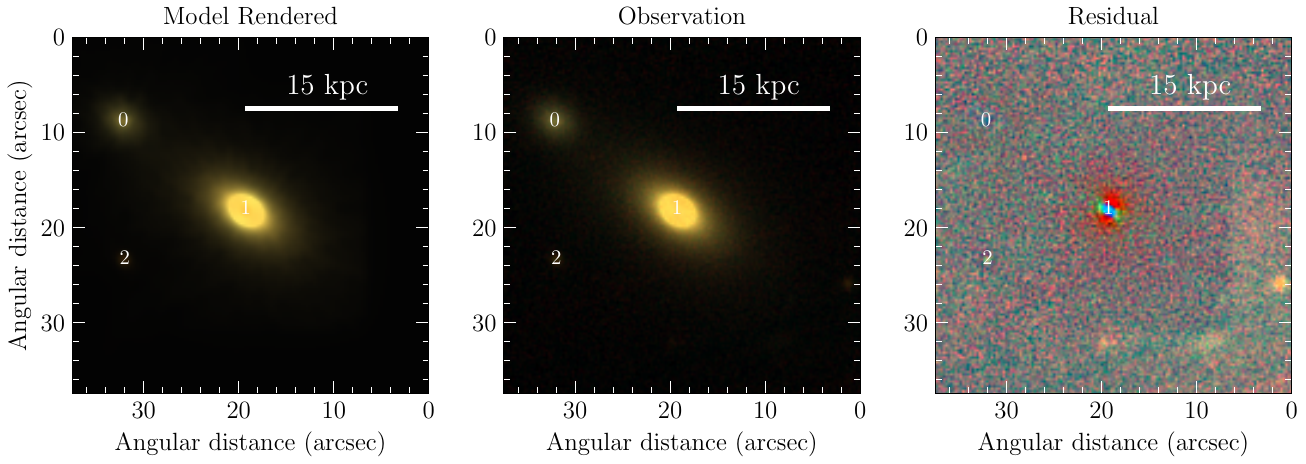}{0.7\textwidth}{Pan-STARRS imaging ($grizy$)}}
\caption{\textit{Top}: \texttt{Scarlet} scene model from LS imaging. Each labeled source was modeled as a monotonically decreasing profile.  We show the model rendered to match the LS imaging (left), the coadded LS $grz$ image (center), and the residual (right). The scene model consists of 4 extended sources (objects labeled 0 to 3). \textit{Bottom}: Same as above but for PS1 $grizy$ imaging. The scene model consists of 3 extended sources (objects labeled 0 to 2).}
\label{fig:Scarletmodel}
\end{figure*}


\begin{figure*}[htbp!]
\centering
\includegraphics[width=0.8\textwidth]{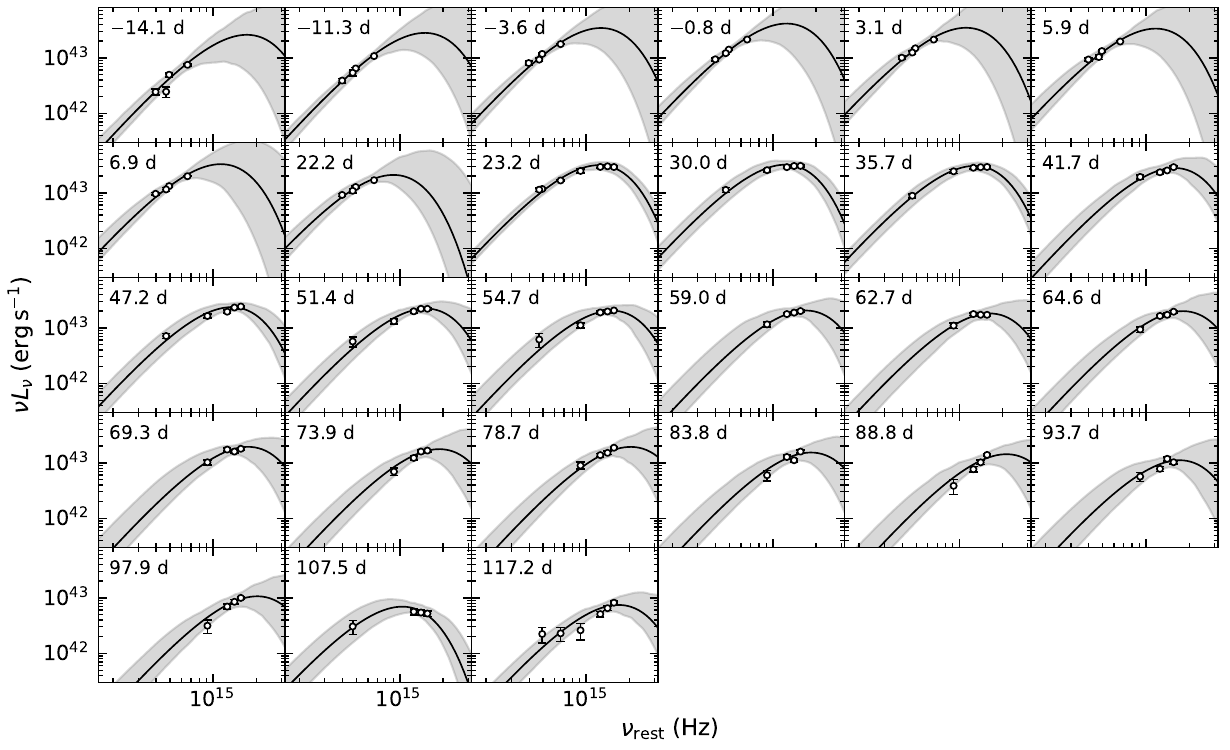}
\caption{UV and optical SED of AT2024tvd, overplotted with the best-fit blackbody models (see Section~\ref{subsec:bbfit}).\label{fig:bbfits}}
\end{figure*}

\begin{figure}[htbp!]
\gridline{\fig{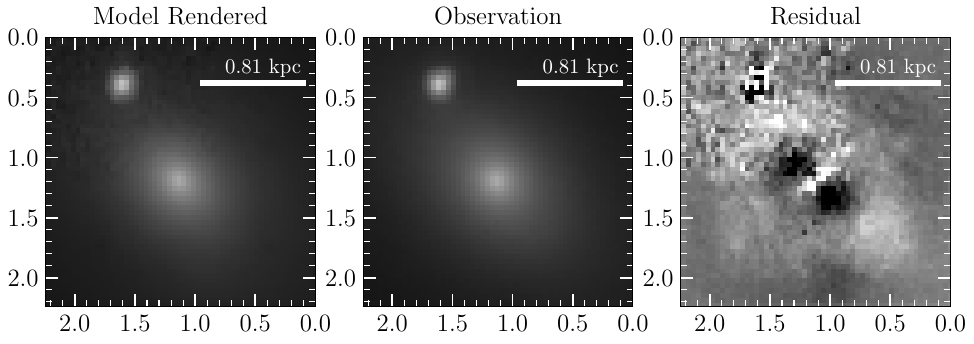}{0.54\textwidth}{Parametric galaxy profile}} 
\gridline{\fig{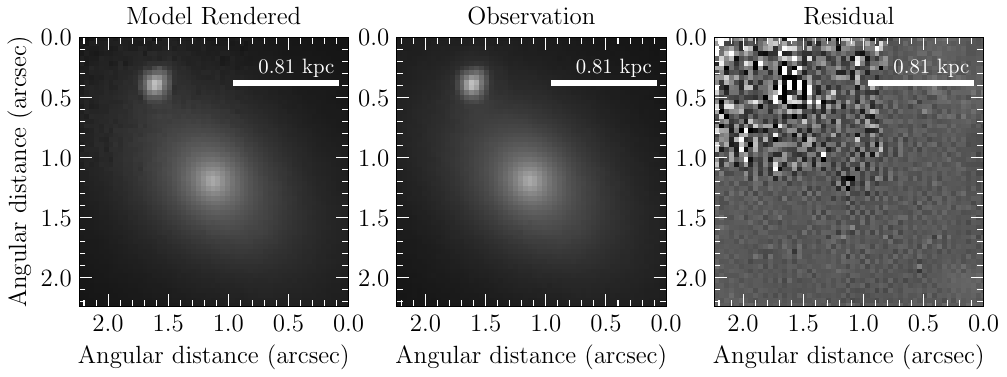}{0.54\textwidth}{Non-parametric galaxy profile}}
\caption{\texttt{Scarlet} scene models of the HST F625W observation. We show 2.2$^{\prime \prime}$ cutouts of the model rendered to match the HST imaging (left), the HST image (center), and the residual (right) for two sets of host galaxy models: 
a Spergel profile (top) and a non-parametric, monotonically decreasing galaxy profile (below). In each case the TDE was modeled as a single PSF, and we do not see any additional extended emission centered on the TDE or any tidal tails associated with it. We do see that the Spergel profile cannot fully describe the bipolar structures at the galaxy center. Some noise is introduced by the low S/N PSF model both around the TDE and the galaxy nucleus.}
\label{fig:HSTScarletmodel}
\end{figure}

\begin{figure}[htbp!]
\gridline{\fig{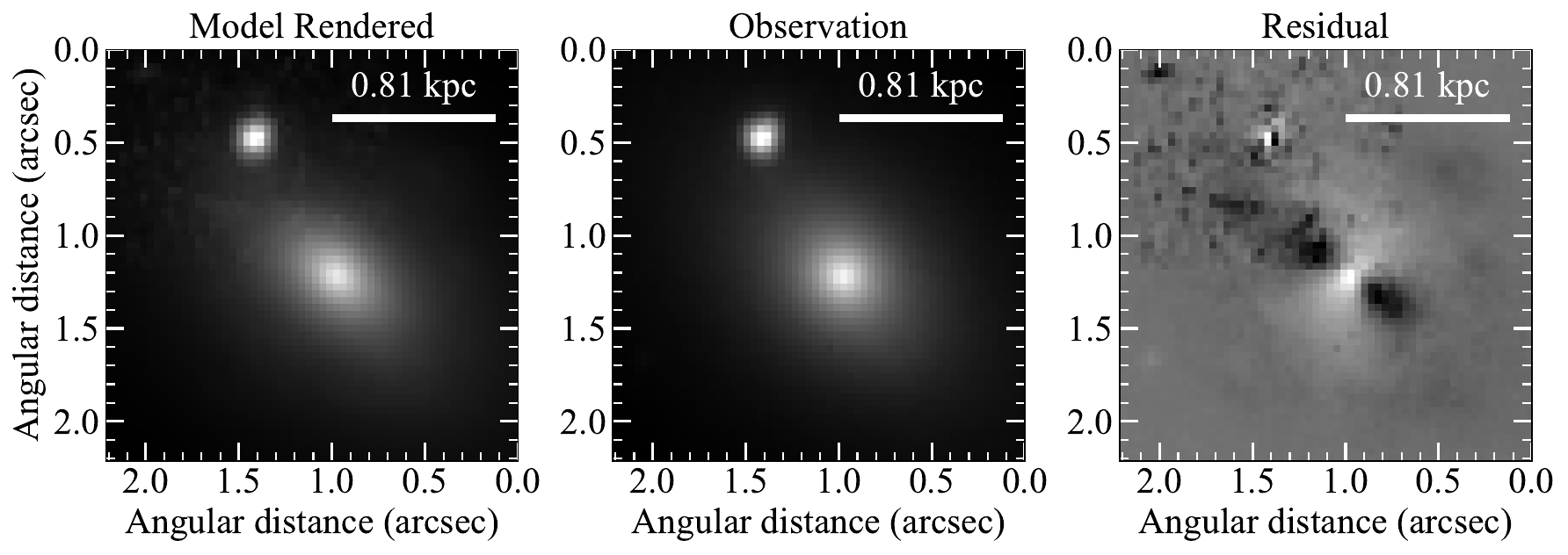}{0.55\textwidth}{Single S\'ersic profile}} 
\gridline{\fig{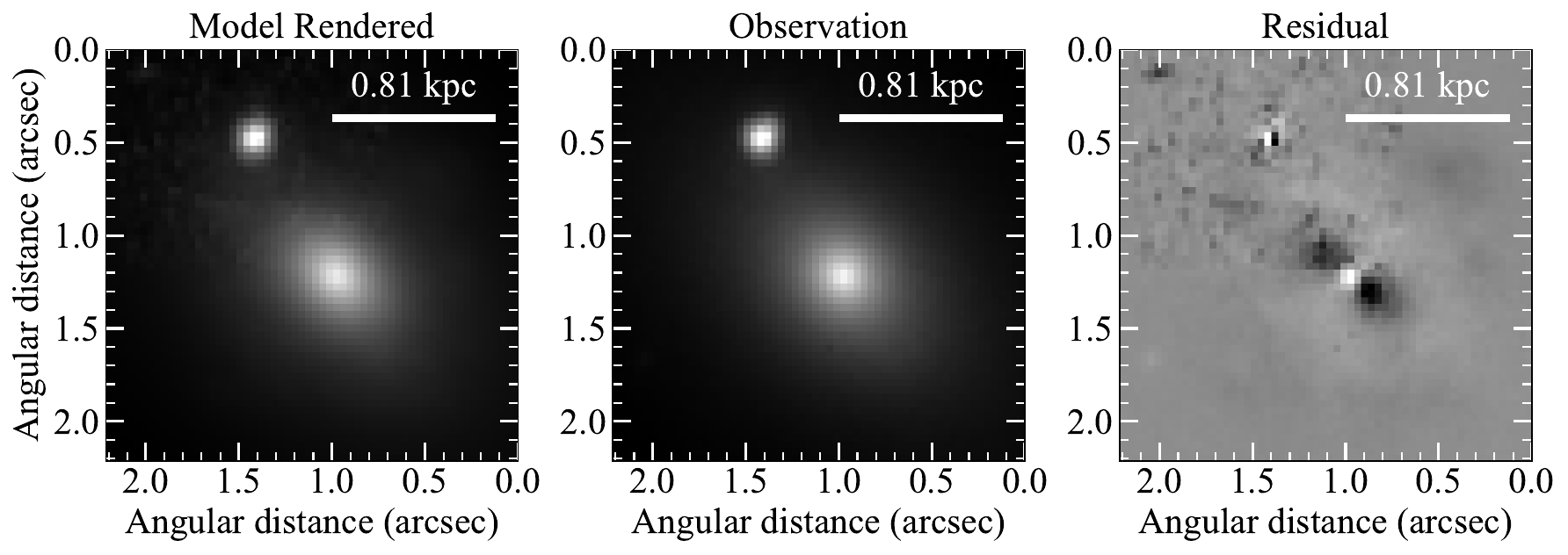}{0.55\textwidth}{Double S\'ersic profile}}
\caption{\texttt{GALFIT} scene models of the HST F625W observation. We show 2.2$^{\prime \prime}$ cutouts of the model rendered to match the HST imaging (left), the HST image (center), and the residual (right) for two sets of host galaxy models: 
a single S\'ersic profile (top) and a double S\'ersic profile (below). In each case the TDE was modeled as a single PSF, and we do not see any additional extended emission centered on the TDE or any tidal tails associated with it. Some noise is introduced by the PSF model.}
\label{fig:HSTgalfitmodel}
\end{figure}

Figure~\ref{fig:Scarletmodel} shows the \texttt{Scarlet} scene model of the pre-TDE LS and PS1 images. 
Figure~\ref{fig:bbfits} shows the blackbody fits on the TDE UV and optical photometry. 
Figure~\ref{fig:HSTScarletmodel} and Figure~\ref{fig:HSTgalfitmodel} show the \texttt{Scarlet} and \texttt{GALFIT} scene models of the HST image, respectively.

\section{Additional Data Tables} \label{sec:table}

We present the UV and optical photometry in Table~\ref{tab:phot}, and the XRT information in Table~\ref{tab:xrt}.

\begin{deluxetable}{ccccc}
\tablecaption{UV and optical photometry of AT2024tvd. \label{tab:phot}}
\tabletypesize{\footnotesize}
	\tablehead{
		\colhead{MJD} &
		\colhead{Instrument} &
		\colhead{Filter} &
		\colhead{$f_\nu$ ($\mu$Jy)} &
        \colhead{$m$ (mag)} 
	}
	\startdata
60546.2347 & ZTF & $r$ & $32.28 \pm 9.84$ & $20.13 \pm 0.33$ \\
60546.2553 & ZTF & $g$ & $32.31 \pm 5.26$ & $20.13 \pm 0.18$ \\
60546.2581 & ATLAS & $o$ & $31.56 \pm 30.86$ & $<18.98$ \\
60551.1931 & ZTF & $i$ & $85.24 \pm 20.91$ & $19.07 \pm 0.27$ \\
60553.2817 & ATLAS & $o$ & $157.62 \pm 22.90$ & $18.41 \pm 0.16$ \\
60577.2489 & ATLAS & $c$ & $571.45 \pm 22.51$ & $17.01 \pm 0.04$ \\
60593.2299 & UVOT & $uvw1$ & $512.76 \pm 28.60$ & $17.13 \pm 0.06$ \\
60593.2319 & UVOT & $U$ & $596.42 \pm 44.18$ & $16.96 \pm 0.08$ \\
60593.2360 & UVOT & $uvw2$ & $409.76 \pm 19.24$ & $17.37 \pm 0.05$ \\
60593.2434 & UVOT & $uvm2$ & $467.67 \pm 21.09$ & $17.23 \pm 0.05$ \\
60676.4629 & UVOT & $U$ & $34.44 \pm 22.55$ & $<19.32$ \\
	\enddata
	\tablecomments{$f_\nu$ is flux density corrected for Galactic extinction, \ad{and $m$ is the corresponding AB magnitude. A subset of the observations is shown for illustration.}
	(This table is available in its entirety in machine-readable form in the online article.)}
\end{deluxetable}

\begin{deluxetable}{crccccccc}
\tablecaption{Swift/XRT observations of AT2024tvd.\label{tab:xrt}}
\tablehead{
    \colhead{obsID}
    & \colhead{MJD} 
    & \colhead{$\delta t$} 
    & \colhead{Exp.}
    & \colhead{Net count rate}
    & \colhead{$f_{\rm X}$}
    & \colhead{$L_{\rm X}$}
    & \colhead{\ad{log$T_{\rm max}$}}
    & \colhead{\ad{$f_{\rm sc}$}}
    \\
    \colhead{}
    & \colhead{}
    & \colhead{(d)}
    & \colhead{(s)} 
    & \colhead{(count\,s$^{-1}$)} 
    & \colhead{($10^{-13}\,{\rm erg\,s^{-1}\,cm^{-2}}$)}
    & \colhead{($10^{42}\,{\rm erg\,s^{-1}}$)}
    & \colhead{(K)}
    & \colhead{}
    }
\startdata
16860001 & $60593.2385\pm0.0094$ & 23.20 & 1626 & $0.0027\pm0.0018$ & $0.83 \pm 0.56$ & $0.39 \pm 0.26$ & \multirow{3}{*}{\ad{$6.37_{-0.06}^{+0.07}$}} &  \\
16860002 & $60600.1805\pm0.3612$ & 29.84 & 1932 & $0.0042\pm0.0019$ & $1.28 \pm 0.58$ & $0.61 \pm 0.27$ &  &  \\
16860003 & $60606.2767\pm0.0681$ & 35.67 & 1808 & $0.0107\pm0.0028$ & $3.23 \pm 0.85$ & $1.53 \pm 0.40$ &  &  \\
\hline
16860004 & $60612.5389\pm0.3638$ & 41.67 & 1629 & $0.0253\pm0.0044$ & $8.47 \pm 1.47$ & $4.01 \pm 0.70$ & \multirow{2}{*}{\ad{$6.14_{-0.03}^{+0.04}$}} &  \\
16860005 & $60618.1951\pm0.2327$ & 47.08 & 2692 & $0.0198\pm0.0030$ & $6.63 \pm 1.01$ & $3.15 \pm 0.48$ &  &  \\
\hline
16860006 & $60622.7140\pm0.1672$ & 51.40 & 2477 & $0.0360\pm0.0041$ & $13.36 \pm 1.51$ & $6.34 \pm 0.71$ & \ad{$6.02_{-0.03}^{+0.04}$} &  \\
\hline
16860007 & $60626.1887\pm0.1370$ & 54.73 & 1994 & $0.0111\pm0.0028$ & $4.81 \pm 1.23$ & $2.28 \pm 0.58$ & \multirow{2}{*}{\ad{$6.14_{-0.04}^{+0.05}$}} &  \\
16860008 & $60630.8945\pm0.3027$ & 59.23 & 2500 & $0.0156\pm0.0029$ & $6.79 \pm 1.24$ & $3.22 \pm 0.59$ &  &  \\
\hline
\multirow{2}{*}{16860009} & $60634.3642\pm0.0094$ & 62.55 & 1630 & $0.0113\pm0.0031$ & $5.58 \pm 1.51$ & $2.64 \pm 0.72$ & \multirow{2}{*}{\ad{$6.00_{-0.05}^{+0.05}$}} &  \\
 & $60634.6292\pm0.0062$ & 62.81 & 1073 & $0.0298\pm0.0058$ & $14.72 \pm 2.88$ & $6.98 \pm 1.37$ &  &  \\
\hline
16860010 & $60636.5225\pm0.0094$ & 64.62 & 1626 & $0.0472\pm0.0058$ & $16.90 \pm 2.07$ & $8.01 \pm 0.98$ & \ad{$6.11_{-0.06}^{+0.06}$} & \ad{$0.04_{-0.02}^{+0.02}$} \\
\hline
16860011 & $60641.4278\pm0.0098$ & 69.31 & 1678 & $0.0162\pm0.0036$ & $9.39 \pm 2.08$ & $4.45 \pm 0.98$ & \ad{$6.08_{-0.09}^{+0.10}$} & \ad{$0.03_{-0.02}^{+0.03}$} \\
\hline
\multirow{2}{*}{16860012} & $60646.1987\pm0.0082$ & 73.88 & 1424 & $0.0724\pm0.0076$ & $24.37 \pm 2.54$ & $11.55 \pm 1.21$ & \multirow{2}{*}{\ad{$6.09_{-0.07}^{+0.09}$}} & \multirow{2}{*}{\ad{$0.08_{-0.02}^{+0.03}$}} \\
 & $60646.2666\pm0.0029$ & 73.94 & 494 & $0.0411\pm0.0103$ & $13.84 \pm 3.47$ & $6.56 \pm 1.65$ &  &  \\
\hline
16860013 & $60651.2378\pm0.0094$ & 78.70 & 1621 & $0.0860\pm0.0077$ & $42.87 \pm 3.85$ & $20.32 \pm 1.82$ & \ad{$6.17_{-0.07}^{+0.07}$} & \ad{$0.14_{-0.03}^{+0.04}$} \\
\hline
16860014 & $60656.6010\pm0.3328$ & 83.83 & 1574 & $0.1004\pm0.0084$ & $37.61 \pm 3.13$ & $17.83 \pm 1.49$ & \ad{$6.05_{-0.07}^{+0.09}$} & \ad{$0.09_{-0.02}^{+0.02}$} \\
\hline
16860015 & $60661.7668\pm0.0096$ & 88.78 & 1653 & $0.0489\pm0.0058$ & $16.28 \pm 1.94$ & $7.72 \pm 0.92$ & \ad{$6.29_{-0.07}^{+0.08}$} & \ad{$0.08_{-0.04}^{+0.05}$} \\
\hline
16860016 & $60666.9255\pm0.0093$ & 93.71 & 1598 & $0.1236\pm0.0092$ & $39.69 \pm 2.94$ & $18.82 \pm 1.40$ & \ad{$6.22_{-0.05}^{+0.05}$} & \ad{$0.12_{-0.03}^{+0.03}$} \\
\hline
\multirow{2}{*}{16860017} & $60671.2398\pm0.0085$ & 97.84 & 1464 & $0.0560\pm0.0066$ & $29.04 \pm 3.42$ & $13.77 \pm 1.62$ & \multirow{2}{*}{\ad{$6.05_{-0.15}^{+0.13}$}} & \multirow{2}{*}{\ad{$0.18_{-0.09}^{+0.06}$}} \\
 & $60671.3045\pm0.0032$ & 97.90 & 548 & $0.1246\pm0.0161$ & $64.55 \pm 8.34$ & $30.60 \pm 3.95$ &  &  \\
\hline
16860018 & $60676.4631\pm0.1961$ & 102.84 & 2060 & $0.0796\pm0.0065$ & $25.47 \pm 2.09$ & $12.08 \pm 0.99$ & \ad{$6.26_{-0.05}^{+0.06}$} & \ad{$0.10_{-0.03}^{+0.03}$} \\
\hline
16860019 & $60681.3025\pm0.0093$ & 107.47 & 1596 & $0.0821\pm0.0075$ & $29.66 \pm 2.73$ & $14.06 \pm 1.29$ & \ad{$6.11_{-0.06}^{+0.07}$} & \ad{$0.11_{-0.03}^{+0.03}$} \\
\hline
\multirow{3}{*}{16860020} & $60691.3542\pm0.0063$ & 117.09 & 1080 & $0.1264\pm0.0113$ & $49.14 \pm 4.41$ & $23.30 \pm 2.09$ & \multirow{3}{*}{\ad{$6.12_{-0.10}^{+0.10}$}} & \multirow{3}{*}{\ad{$0.25_{-0.05}^{+0.06}$}} \\
 & $60691.4226\pm0.0025$ & 117.16 & 425 & $0.0551\pm0.0127$ & $21.42 \pm 4.95$ & $10.16 \pm 2.35$ &  &  \\
 & $60691.6180\pm0.0028$ & 117.34 & 489 & $0.1582\pm0.0191$ & $61.48 \pm 7.42$ & $29.15 \pm 3.52$ &  &  \\
 \enddata
\tablecomments{Net count rate, observed flux $f_{\rm X}$, and the observed luminosity $L_{\rm X}$ are given in 0.3--10\,keV. }
\end{deluxetable}

\bibliography{tde}{}
\bibliographystyle{aasjournalv7}

\end{document}